\documentclass[journal=jctcce]{achemso}

\usepackage{graphicx}
\usepackage{amssymb}
\usepackage{amsmath}
\usepackage{amsbsy}
\usepackage{multirow}
\usepackage{hyperref}
\usepackage{bm}
\usepackage{color}
\usepackage{bbold}
\usepackage[english]{babel}
\makeatletter
\@namedef{l@en}{\l@english}
\makeatother

\usepackage{epstopdf}
\usepackage{ulem}
\usepackage{verbatim}
\usepackage{afterpage}
\usepackage{placeins}
\usepackage{epstopdf}
\hypersetup{
    colorlinks=true,
    linkcolor=blue,
    filecolor=magenta,      
    urlcolor=cyan,
    pdftitle={BOCSv5},
    pdfpagemode=FullScreen,
    }
\DeclareMathAlphabet\mathbfcal{OMS}{cmsy}{b}{n}

\newcommand{\tsup}[1]{\textsuperscript{#1}}

\newcommand{\llangle}{\left\langle}
\newcommand{\rrangle}{\right\rangle}

\newcommand{\tallA}{\vphantom{\hat{A}}}

\newcommand{\inv}{^{-1}}

\newcommand{\half}{\frac{1}{2}}

\newcommand{\dd}{{\rm d}}
\newcommand{\ptl}{\partial}
\newcommand{\kB}{k_B}
\newcommand{\kT}{\kB T}
\newcommand{\del}{{\bm{\nabla}}}
\newcommand{\dI}{\del_I}
\newcommand{\dJ}{\del_J}
\newcommand{\al}{\alpha}
\newcommand{\bt}{\beta}
\newcommand{\de}{\delta}

\newcommand{\ga}{\gamma}

\newcommand{\la}{\lambda}

\newcommand{\te}{\theta}

\newcommand{\zt}{\zeta}

\newcommand{\cDn}{{\mathcal{D}}_n}
\newcommand{\cDN}{{\mathcal{D}}_N}
\newcommand{\cS}{{\mathcal{S}}}
\newcommand{\cSI}{{\mathcal{S}}_I}

\newcommand{\br}{{\bf r}}
\newcommand{\bri}{\br_i}
\newcommand{\bR}{{\mathbf R}}
\newcommand{\bRI}{\bR_I}

\newcommand{\RIJ}{R_{IJ}}

\newcommand{\hbR}{\widehat\bR}
\newcommand{\hbRJK}{\hbR_{JK}}
\newcommand{\hbRIJ}{\hbR_{IJ}}

\newcommand{\invnm}{nm$^{-3}$}

\newcommand{\bM}{{\mathbf M}}
\newcommand{\dMrR}{\de\left(\bM(\br) - \bR\right)}

\newcommand{\cIi}{c_{Ii}}

\newcommand{\dr}{\dd\br}

\newcommand{\intVr}{\int_{\cDn(V)} \!\!\!\!\!\!\!\!\dr \:}

\newcommand{\sumin}{\sum_{i=1}^n}
\newcommand{\sumIN}{\sum_{I=1}^N}

\newcommand{\pr}{p_{\rm r}}

\newcommand{\pR}{p_{\rm R}}

\newcommand{\zR}{z_{\rm R}}

\newcommand{\rhoL}{\rho_{\rm L}}

\newcommand{\UR}{U_{\rm R}}
\newcommand{\Ub}{U_{\rm b}}
\newcommand{\Unb}{U_{\rm nb}}
\newcommand{\UV}{U_{\rm V}}

\newcommand{\Upair}{U_{\rm pair}}
\newcommand{\Uld}{U_{\rm LD}}
\newcommand{\Usg}{U_{\rm SG}}

\newcommand{\Utwo}{U_2}
\newcommand{\Urho}{U_\rho}
\newcommand{\Udel}{U_\del}

\newcommand{\Uzt}{U_\zt}

\newcommand{\rhoI}{\rho_I}
\newcommand{\rhoJ}{\rho_J}
\newcommand{\wbar}{\overline w}
\newcommand{\wb}{\overline{w}}

\newcommand{\bff}{{\bf f}}
\newcommand{\bfi}{\bff_i}
\newcommand{\bfI}{\bff_I}
\newcommand{\bF}{{\bf F}}
\newcommand{\bFI}{\bF_I}
\newcommand{\FV}{F_{\rm V}}

\newcommand{\bFInb}{\bF_{I;\rm nb}}
\newcommand{\bFIJnb}{\bF_{IJ;\rm nb}}
\newcommand{\bFIJsg}{\bF_{IJ;\rm SG}}

\newcommand{\Ftwo}{F_2}
\newcommand{\Frho}{F_\rho}
\newcommand{\Fdel}{F_\del}

\newcommand{\bA}{{\bf A}}

\newcommand{\bAIJ}{\bA_{IJ}}

\newcommand{\Fzt}{F_\zt}
\newcommand{\fzd}{f_{\zt d}}
\newcommand{\phizd}{\phi_{\zt d}}
\newcommand{\fdeld}{f_{\del d}}
\newcommand{\phideld}{\phi_{\del d}}

\newcommand{\cG}{{\mathbfcal{G}}}

\newcommand{\cGIzd}{{\mathbfcal{G}}_{I;\zt d}}
\newcommand{\cGID}{{\mathbfcal{G}}_{I;D}}
\newcommand{\cGIDp}{{\mathbfcal{G}}_{I;D'}}
\newcommand{\cGD}{{\mathbfcal{G}}_{D}}

\newcommand{\bD}{b_{D}}
\newcommand{\phiD}{\phi_{D}}
\newcommand{\phiDp}{\phi_{D'}}
\newcommand{\GDDp}{G_{DD'}}

\newcommand{\Tref}{T_0}
\newcommand{\Pref}{P_0}

\newcommand{\rc}{r_{\rm c}}
\newcommand{\rcut}{r_{\rm cut}}

\newcommand{\pmol}{mol$^{-1}$}

\newcommand{\tUtwo}{\widetilde{U}_2}
\newcommand{\tFtwo}{\widetilde{F}_2}

\newcommand{\Pf}{P_{\rm f}}
\newcommand{\Pb}{P_{\rm b}}

\newcommand{\rdel}{r_\del}
\newcommand{\rhoLd}{\rho_{\rm L \del}}

\newcommand{\rhomax}{\rho_{\rm max}}
\newcommand{\tUtt}{\widetilde{U}_{2t}}

\newcommand{\brho}{\overline{\rho}}
\newcommand{\brhoL}{\brho_{\rm L}}

\newif\ifShowFig
\ShowFigtrue
\newif\ifShowBigFig
\ShowBigFigfalse

\graphicspath{{./Figures/},{./}}

\title{Progress toward a better BOCS: Systematic coarse-graining with local density potentials}
\author{Maria C. Lesniewski}
\author{Michael R. DeLyser}
\author{W.~G.~Noid}
\email{wgn1@psu.edu}
\affiliation{Department of Chemistry,  The Pennsylvania State University,  University Park, Pennsylvania 16802, USA}
\date{June 15th, 2026}

\begin{document}		

\begin{abstract}
We describe version 5.0 of the Bottom-up Open-source Coarse-graining Software (BOCS) package.
BOCS employs the force-matching variational principle to parameterize potentials for coarse-grained (CG) models directly from atomically detailed simulations.
BOCS version 5.0 significantly extends previous versions by treating potentials that depend upon the local density (LD) around each particle, as well as  potentials that depend upon the square gradient (SG) of this local density.
We also describe a new package, PKG-BOCS, for simulating these potentials in LAMMPS. 
This software treats complex molecular topologies and provides considerable flexibility for defining the local density, as well as the LD and SG potentials.
We present numerical calculations that provide physical insight into these potentials and demonstrate the accuracy of our implementation.
Finally, we demonstrate that LD potentials can significantly improve the structural fidelity, thermodynamic properties, and transferability of CG models for water.
\end{abstract}

\maketitle

\section{Introduction}
By representing systems in reduced detail, coarse-grained (CG) models provide the necessary computational efficiency for simulating length- and time-scales that cannot be easily addressed with atomically detailed models.\cite{Deserno:2009wd,Peter:2010zc,Guenza:2018tr,Gartner:2019}
Coarse-graining also provides conceptual advantages by eliminating unnecessary details and focusing attention on essential features.\cite{Muller:2006ij,Schmid:2009cr,Giulini:2021wn}
There exist many approaches for developing CG models with varying advantages and limitations.\cite{Noid:2013cr,Ingolfsson:2014rz,borges-araujo_pragmatic_2023}
Here we focus on ``bottom-up'' CG models that are parameterized with information from simulations of all-atom (AA) or other high resolution models.\cite{jin_bottom-up_2022,Noid:2023a}
In principle, statistical mechanics provides a framework for parameterizing bottom-up models that exactly reproduce all structural and thermodynamic properties of the underlying AA model that can be observed at the resolution of the CG model.\cite{Likos:2001kx,Dunn:2016bb}
In practice, though, this exact coarse-graining procedure is not analytically tractable and must be computationally approximated.

Several open-source software packages implement bottom-up methods for approximating this exact coarse-graining procedure, including  the versatile object-oriented toolkit for coarse-graining applications (VOTCA),\cite{Ruhle:2009wx,Baumeier2024} IBIsCO\cite{Karimi-Varzaneh:2011fk}, MagiC \cite{Mirzoev:2013fk,mirzoev_magic_2019}, the bottom-up open-source coarse-graining software (BOCS) package,\cite{dunn_bocs:_2018}  and the Open multiscale coarse-graining\cite{Noid:2008a} (OpenMSCG) package.\cite{Lu:2010dp,peng2023openmscg}
VOTCA, IBIsCO, MagiC, and OpenMSCG implement structure-based methods that iteratively refine CG potentials to reproduce target distribution functions describing the mapped ensemble, i.e., the ensemble obtained by mapping the equilibrium ensemble for the underlying AA model to the CG representation.\cite{Lyubartsev:1995,MullerPlathe:2002,Brini:2013wn}
If they converge, these methods maximize the likelihood that the CG model will reproduce the mapped ensemble.\cite{Shell:2008il,Chaimovich:2011fk,Noid:2013uq}
Conversely, VOTCA, BOCS, and OpenMSCG implement a force-matching\cite{Ercolessi:1994,Chorin:2003,Izvekov:2005d,Izvekov:2005e} variational principle\cite{Noid:2008a,Noid:2008b} that directly (i.e., noniteratively) parameterizes the CG potential to optimally approximate the configuration-dependence of the exact CG potential when evaluated over the mapped ensemble. 
This force-matching (FM) variational approach is closely related\cite{noid_rigorous_2024} to the score-matching method\cite{hyvarinen_estimation_2005} for parameterizing machine learning  methods.\cite{song_how_2021}

BOCS complements other force-matching implementations by providing several unique capabilities.
For instance, BOCS employs a generalized-Yvon-Born-Green (g-YBG) framework\cite{Noid:2007a,Rudzinski:2015bb} to perform force-matching without forces by directly determining the CG potential from structural information.\cite{Mullinax:2009b,Mullinax:2010}
This g-YBG framework clarifies and quantifies the relationship between the mapped ensemble and the CG potential.\cite{Rudzinski:2012vn,Rudzinski:2015bb} 
Additionally, BOCS implements a simple framework for improving the transferability of CG potentials across multiple systems and thermodynamic conditions.\cite{Mullinax:2009a}
Moreover, BOCS includes software for parameterizing configuration-independent potentials of the volume.\cite{Das:2010,Dunn:2015wn} 
CG models with these volume potentials can quantitatively reproduce the AA pressure-density equation of state across a very wide range of thermodynamic conditions.\cite{Dunn:2016aa,lesniewski_insight_2024}  
Volume potentials can also reasonably describe density fluctuations in metal organic frameworks.\cite{alvares_force_2024} 
However, volume potentials cannot be employed to model liquid-vapor interfaces or other highly inhomogeneous systems.

A growing number of studies have modeled such inhomogeneous systems by supplementing conventional pair potentials with potentials of the local density around each molecule. 
By defining the local density with additive contributions from surrounding molecules, local density (LD) potentials generate short-ranged pair forces that are sensitive to the local environment. 
While conceptually similar to the embedded atom model for metals,\cite{daw_embedded-atom_1984,daw_embedded-atom_1993} LD potentials for soft materials were first introduced in the many-body dissipative particle dynamics model developed by Pagonabarraga and Frenkel.\cite{Pagonabarraga:2001aa}
Wolynes and coworkers independently introduced LD potentials for modeling water-mediated interactions in protein structure prediction.\cite{Papoian:2004}
More recently, bottom-up models have employed LD potentials to describe liquids,\cite{Allen:2008,Wagner:2017xx,DeLyser:2017,Delyser:2019ld,Delyser:2020int,Szukalo:2023,Dutta:2026wn,dutta_coarse-graining_2026} liquid mixtures,\cite{Allen:2009,sanyal_transferable_2018,rosenberger_transferability_2019} polymers,\cite{Sanyal:2016md,Agrawal:2016md,shahidi_coarse-graining_2020,berressem_ultra-coarse-graining_2021} and even explosive materials.\cite{Moore:2016wn,izvekov_bottom-up_2022} 
These LD potentials can significantly improve the pressure-density equation of state for bottom-up models\cite{DeLyser:2017,Delyser:2019ld} and provide outstanding transferability between bulk and interfacial environments.\cite{Delyser:2020int} 
LD potentials also reasonably reproduce the cooperative many-body interactions that govern hydrophobic phenomena.\cite{Lesniewski:2026a}
Moreover, following the analogy to classical density functional theory,\cite{Evans:1979fk,Pagonabarraga:2001aa} we have introduced a new class of interaction potentials that depend upon both the local density and the square of its gradient.\cite{Delyser:2022wn}
The combination of these LD and square gradient (SG) potentials can model interfacial density profiles very accurately, while still employing only pair-additive forces.

Here, we describe a new release of BOCS for parameterizing LD and SG potentials via variational force-matching.
We also describe a new version of the PKG-BOCS package for simulating these potentials in the LAMMPS molecular dynamics engine.\cite{Plimpton:1995,thompson_lammps_2022}
This software allows for multiple different types of particles and corresponding local densities.\cite{Dutta:2026wn} 
It allows users to define each type of local density with a unique length-scale and with one of five different weighting functions.
The software implements simple functional forms and also more flexible spline-based representations for LD and SG potentials. 
Additionally, this software is compatible with complex CG topologies, while properly accounting for intramolecular contributions and exclusions. 
The updated package includes detailed documentation and tutorials that describe and illustrate the software. 
Most importantly, the software is accurate and robust, as it can quantitatively determine tens of thousands of parameters for systems with hundreds of molecules that are governed by bond, angle, dihedral, pair, LD, and SG potentials. 
We hope that this new software will provide a useful complement to existing packages for parameterizing and simulating LD potentials.\cite{sanyal_transferable_2018,Sanyal:2016md,peng2023openmscg}

The remainder of this manuscript is organized as follows:
The theory section describes the relevant background for LD and SG potentials. 
The methods section describes the computational workflow for parameterizing and simulating these potentials. 
This section also summarizes the details of numerical calculations that illustrate this new software. 
The results section presents numerical calculations that illustrate the accuracy and robustness of the new software, while providing physical insight into LD and SG potentials.
The concluding section provides closing comments and suggests promising directions for future work. 
The appendix provides further analysis of LD and SG potentials, while the supporting information (SI) provides additional computational details.

\section{Theory}
\label{sec-Theory}
\subsection{Mapped ensemble}
\label{subsec-MappedEnsemble}
Bottom-up approaches parameterize CG models based upon information from a higher resolution model.\cite{jin_bottom-up_2022,Noid:2023a}
We shall refer to this high resolution model as an all-atom (AA) model for simplicity, but the present framework applies more generally to classical particle-based models.
We denote the configuration of the AA model by $\br = (\br_1, \ldots, \br_n)$,  where $\bri$ denotes the Cartesian coordinates of atom $i$.
We assume that the system is at equilibrium, such that the AA model samples each configuration according to the canonical probability density, $\pr(\br) \propto \exp[-\bt u(\br)]$, where $\bt = 1/\kT$ is the inverse temperature and $u(\br)$ is the corresponding potential.\cite{Tuckerman:2013}
While we focus on the canonical ensemble, this framework can be generalized to other ensembles.\cite{Das:2010,Dunn:2015wn}

We employ a mapping, $\bM: \br\to\bR=\bM(\br)$, to define the CG representation, $\bR = (\bR_1, \ldots, \bR_N)$, of each AA configuration.\cite{Noid:2008a}
This mapping defines the coordinates of each site, $I$, as a convex combination of atomic coordinates, i.e., $\bRI = \sumin \cIi \bri$.\cite{kidder_analysis_2024}
In particular, we consider ``non-overlapping'' maps for which each atom contributes to the definition of at most one CG site.
We define a restricted partition function for the total Boltzmann weight associated with each CG configuration, $\bR$:
\begin{equation}
\label{def-zR}
\zR(\bR) \equiv \intVr \exp[-\bt u(\br)] \dMrR	,
\end{equation}	
where $\cDn(V)$ denotes the set of AA configurations that are accessible when the system has a volume, $V$.\cite{kidder_analysis_2024}
This restricted partition function determines the mapped probability density, $\pR(\bR) \propto \zR(\bR)$, as well as the exact CG potential:
\begin{equation}
\label{def-W}
W(\bR) \equiv - \kT \ln \left[\zR(\bR) / V_0^{n-N} \right]	,
\end{equation}
where $V_0$ is a constant reference volume.\cite{Noid:2008a}
If $W$ is known as a function of both configuration and also thermodynamic conditions, then it is possible to quantitatively reproduce all structural and thermodynamic properties that are observable at the resolution of the CG mapping.\cite{Likos:2001kx,Dunn:2016bb}

\subsection{Approximate CG potential}
\label{subsec-ApproxPotl}
BOCS approximates $W$ with CG potentials of the form
\begin{equation}
\label{def-U}
U(\bR,V) \equiv \Ub(\bR) + \Unb(\bR) + \UV(V)		.
\end{equation}
Here $\Ub(\bR)$ is a conventional bonded potential, $\Unb(\bR)$ is the non-bonded potential, and $\UV(V)$ is a volume potential.

The bonded and volume potentials in Eq.~\eqref{def-U} have not changed in this latest version of BOCS. 
The bonded potential includes terms governing  
bond, angle, and torsional degrees of freedom.
These terms can be defined with simple analytic forms (e.g., harmonic bond and angle potentials) or with more flexible basis functions (e.g., linear or cubic spline functions).
The volume potential does not generate forces on any particles but directly contributes to the internal pressure via $\FV(V) = - \dd\UV(V)/\dd V$.
BOCS can represent this volume force with a two parameter functional form\cite{Das:2010} or with more flexible basis functions.\cite{Dunn:2015wn,Dunn:2016aa,lesniewski_insight_2024}

BOCS now treats three different classes of non-bonded interactions:
\begin{equation}
\label{def-Unb}
\Unb(\bR) 
	\equiv	
			\Upair(\bR) + \Uld(\bR) + \Usg(\bR)	
			.
\end{equation}
Here $\Upair$, $\Uld$, and $\Usg$ denote, respectively, contributions from conventional central pair potentials, local density potentials, and also potentials that depend upon the square of the local density gradient.
BOCS allows for complex molecular topologies and many different types of sites.
However, for the sake of simplicity and clarity of presentation, we shall assume that all CG particles are of the same type in the present subsection.

For each CG site, $I$, we define a set, $\cSI$, of sites that interact with $I$ via non-bonded potentials.
This set can include intramolecular non-bonded interactions, while also accounting for intramolecular exclusions. 
The pair contribution to $\Unb$ may be expressed
\begin{equation}
\label{def-Upair}
\Upair(\bR) \equiv \half \sumIN \sum_{J\in\cSI} \Utwo(\RIJ) 	,
\end{equation}
where $\Utwo$ is a central pair potential and $\RIJ$ is the distance between CG sites $I$ and $J$.

Previous releases of BOCS did not treat $\Uld$ and $\Usg$. 
These terms both depend upon the instantaneous local density, $\rhoI$, around each particle, $I$.
We define $\rhoI$ as a sum of contributions from neighboring particles that participate in non-bonded interactions with $I$: 
\begin{equation}
\label{def-rhoI}
\rhoI 
	\equiv 
		\rhoI(\bR) 
	\equiv 
		c + \sum_{J\in \cSI} \wbar(\RIJ)		.
\end{equation}
Here $\wbar(r)$ is a normalized, non-increasing weighting function that vanishes for all $r \ge \rc$, while $c$ is a configuration-independent constant that can be ignored, $c = 0$, or used to define the ``self-contribution,'' $c = \wbar(0)$.
BOCS allows users to specify the LD length-scale, $\rc$, and to employ five different functional forms for $\wbar$.
We discuss these weighting functions in Section~\ref{SubsubSec-Methods-WeightingFunctions} below and Section~I of the SI. 
We now define 
\begin{eqnarray}
\label{def-Uld}
\Uld(\bR) 
	& \equiv & 	
					\sumIN \Urho(\rhoI)			\\
\label{def-Usg}
\Usg(\bR) 
	& \equiv & 	
					\sumIN \Udel(\rhoI) \left| \dI \rhoI \right|^2			,
\end{eqnarray}
where $\dI\rhoI = \ptl \rhoI / \ptl \bRI$.
Note that $\Udel$ is not a potential and does not have units of energy.
Rather, $\Udel$ is a coefficient that modulates the square gradient contribution as a function of the local density and, in particular, can be defined to act primarily in interfacial regimes.\cite{Delyser:2022wn}  
Because they are defined in terms of the local density, Eq.~\eqref{def-rhoI} ensures that $\Uld$ and $\Usg$ properly treat intramolecular contributions and exclusions.

The total non-bonded force on site $I$ may be expressed
\begin{equation}
\label{def-bFInb}
\bFInb 
	\equiv 
			- \dI \Unb(\bR) 
		= 
			\sum_{J\in\cSI} \bFIJnb(\bR)		,
\end{equation}
where the total non-bonded force between the pair $(I,J$) is
\begin{equation}
\label{def-bFIJnb}
\bFIJnb(\bR)
	\equiv
			\left\{
					\Ftwo(\RIJ)
				+
					\left[ 
							\Frho(\rhoI)
						+
							\Frho(\rhoJ)							
					\right]
					\wbar'(\RIJ)
			\right\}
			\hbRIJ
	+
			\bFIJsg(\bR)
			.
\end{equation}
Here 
$\Ftwo(r) = - \dd \Utwo(r)/\dd r$, 
$\Frho(\rho) = - \dd\Urho(\rho)/\dd \rho$, 
$\wbar'(r) = \dd \wbar(r) / \dd r$, 
$\hbRIJ$ is the unit vector pointing from $J$ to $I$, 
and $\bFIJsg(\bR)$ is the contribution to the pair force from $\Usg$.
The pair and LD potentials generate antisymmetric forces along the vector between each pair. 
Importantly, $\Uld$ modulates these pair forces as a function of the local density around both particles.\cite{Lesniewski:2026a} 
The contribution from $\Usg$ may be expressed\cite{Delyser:2022wn}
\begin{eqnarray}
\nonumber
\bFIJsg(\bR)
	& \equiv & 
			\left[
						\tallA
					\Fdel(\rhoI) \left|\dI \rhoI \right|^2		
				+  
					\Fdel(\rhoJ) \left|\dJ \rhoJ \right|^2		
			\right] 
			\wbar'(\RIJ)
			\hbRIJ
			\\
	&& 
		- 
			2 g(\RIJ) 
						\left[
								\bAIJ
					\cdot 
					\hbRIJ
						\right]
				\hbRIJ
		- 
			2 h(\RIJ) 
				\bAIJ		
					,
\label{def-bFIJsg}
\end{eqnarray}
where 
$\Fdel(\rho) = - \dd \Udel(\rho)/\dd\rho$, 
$h(r) \equiv \wbar'(r) / r$, 
$g(r) = \wbar''(r) - h(r)$,
and 
\begin{equation}
\label{def-bAIJ}
\bAIJ 
	= 
		\Udel(\rhoI) \dI \rhoI	- \Udel(\rhoJ) \dJ \rhoJ
	 .
\end{equation}
While the SG potential generates anti-symmetric pair-additive forces, these forces are not along $\hbRIJ$ because the last term in Eq.~\eqref{def-bFIJsg} introduces contributions along $\dI \rhoI$ and $\dJ \rhoJ$. 
Because $\wbar(r) = 0$ for $r > \rc$, the LD and SG contributions to $\bFIJnb(\bR)$ also vanish when $\RIJ > \rc$.

Equations~\eqref{def-Upair} -- \eqref{def-bAIJ} explicitly consider a single type of particle and, consequently, a single local density.
Appendix 1 generalizes this treatment for CG models with many different types of particles and local densities, along with a corresponding diversity of LD and SG  potentials.

\subsection{Variational force-matching}
BOCS determines the optimal approximate CG potential according to the force-matching variational principle\cite{Chorin:2003,Noid:2008a} that is employed in the multiscale coarse-graining (MS-CG) method,\cite{Izvekov:2005d,Izvekov:2005e} i.e., by minimizing
\begin{equation}
\label{def-chiU}
\chi^2[U] 
	\equiv 
		\llangle 
			\frac{1}{3N} 
			\sumIN 
					\left|	
						\tallA 
						\bFI(\bM(\br))
						-
						\bfI(\br) 
					\right|^2
		\rrangle
		.
\end{equation}  
Here the angular brackets denote an average over the equilibrium ensemble for the AA model, $\bfI(\br)$ indicates the net force on site $I$ in the AA configuration $\br$,\cite{Noid:2008a,kramer_statistically_2023} and $\bFI(\bR) = -\dI U(\bR)$.
Equation~\eqref{def-chiU} can be decomposed\cite{Noid:2008a} into a fixed contribution from the exact CG potential, $W$, and a non-negative contribution comparing the gradients of $W$ and $U$: 
\begin{equation}
\label{eq-chiU-W}
\chi^2[U] 
	= 
		\chi^2[W] 
	+ 
		\int_{\cDN(V)} \!\!\!\!\!\!\!\! \pR(\bR) 
				\frac{1}{3N}
				\sumIN 
					\left|\dI\left( W - U \right)\right|^2	,
\end{equation}
Consequently, $W$ provides the global minimum of $\chi^2[U]$.
Moreover, variations in the CG potential, $U \to U +\de U$, that reduce $\chi^2$ result in a better approximation for the configuration-dependence of $W$.
We now construct a basis set for performing this variational calculation.\cite{Noid:2008b}

Equations~\eqref{def-bFInb} and \eqref{def-bFIJnb} demonstrate that the total nonbonded force, $\bFInb$, on each particle, $I$, depends linearly upon the force functions, $\Ftwo$ and $\Frho$, that are determined by the pair and LD potentials, $\Utwo$ and $\Urho$.
Similarly, the total bonded force from $\Ub$ in Eq.~\eqref{def-U} may also be written as a linear combination of bonded force functions. 
For each bonded, pair, or LD potential, $\Uzt(x)$, we represent the corresponding force function, $\Fzt(x) = - \dd\Uzt(x)/ \dd x$, by a linear combination of basis functions, $\fzd(x)$, with constant coefficients, $\phizd$:
\begin{equation}
\label{def-Fzt}
\Fzt(x) = \sum_d \phizd \fzd(x)	.
\end{equation}
This basis expansion allows us to express the CG force, $\bFI$, on each particle, $I$, as a linear combination of these coefficients,  $\phizd$.
However, this approach does not work for the SG potential because the corresponding force in Eq.~\eqref{def-bFIJsg} depends upon both $\Fdel$ and $\Udel$.
Consequently, we represent $\Udel(x) = \sum_d \phideld \fdeld(x)$, which determines a corresponding representation for $\Fdel(x) = - \sum_d \phideld \fdeld'(x)$.
This allows us to express the total force on each CG site as a linear combination of these coefficients:
\begin{equation}
\label{eq-bFI-cGIzd}
\bFI(\bR;\phi) 
	= 
		\sum_\zt \sum_d \phizd \cGIzd(\bR) 	,
\end{equation}
where $\cGIzd$ specifies the direction of the force on CG site $I$ that arises from the $d$\tsup{th} basis function, $\fzd(x)$, for either the $\zeta$ force function, $\Fzt(x)$, or, in the case of the SG interaction, $\zt = \del$, the coefficient function, $\Udel(x)$.
The coefficient, $\phizd$, then determines the magnitude of this force.
Appendix 1 provides explicit expressions for the vector $\cGIzd(\bR)$, associated with each type of interaction.

If we now consider $\bFI(\bR;\phi)$ to be the $I$\tsup{th} element of a force field, $\bF(\phi) = \{ \bFI(\bR;\phi)\}$, and adopt a `super-index' $D = \zt d$, then Eq.~\eqref{eq-bFI-cGIzd} may re-expressed more simply
\begin{equation}
\label{eq-bF-cGD}
\bF(\phi) 
	= 
		\sum_D \phiD \cGD 	,
\end{equation}
where $\cGD = \{\cGID(\bR)\}$ may be considered a force field basis vector.\cite{Noid:2008b}
The set, $\{\cGD\}$, of basis vectors included in Eq.~\eqref{eq-bF-cGD} specifies a vector space of CG force fields that is defined by the form of the approximate potential, $U$.
A set of coefficients, $\phi \equiv \{\phiD\}$, identifies a particular force field within this subspace and determines a corresponding approximate potential (to within an irrelevant additive constant).

Given this basis set, $\chi^2$ becomes a quadratic form in the potential parameters:
\begin{equation}
\label{eq-chi2-quad}
\chi^2(\phi) 
	\equiv 
			\chi^2[U(\phi)] 
	= 
		\sum_{D,D'} \phiD \GDDp \phiDp - 2 \sum_D \bD \phiD + \chi^2_0	,
\end{equation}
where $\chi^2_0 = \chi^2[0]$ and 
\begin{eqnarray}
\label{def-GDDp}
\GDDp	
		& \equiv &
					\llangle 
						\frac{1}{3N} 
							\sumIN 
										\tallA 
											\cGID(\bM(\br))
											\cdot
											\cGIDp(\bM(\br))
					\rrangle
					\\
\label{def-bD}
\bD
			& \equiv & 
					\llangle 
						\frac{1}{3N} 
							\sumIN 
										\tallA 
											\cGID(\bM(\br))
											\cdot
											\bfI(\br)
					\rrangle		.
\end{eqnarray}
Both $\GDDp$ and $\bD$ are defined as averages over the equilibrium ensemble for the AA model and, thus, may be directly computed from the AA trajectory. 
$\GDDp$ is a structural correlation function describing the correlation between the forces generated by $D$ and $D'$ basis vectors.\cite{Rudzinski:2012vn}
Conversely, $\bD$ may be considered the projection of the atomic force field onto the $D$ basis vector.\cite{Mullinax:2009b}
Because $\bD$ can be related to the mean force of the AA model along the corresponding degree of freedom, it can be computed directly from structural information.\cite{Noid:2007a,Mullinax:2010}
The set of minimizing coefficients, $\phi_* = \{ \phi_{D_*} \}$, that determine the optimal approximate CG potential satisfy the normal system of equations\cite{Noid:2008b}
\begin{equation}
\label{eq-normal}
\sum_{D'} \GDDp \phi_{D'_*} = \bD		.
\end{equation}
BOCS includes several numerical methods for solving Eq.~\eqref{eq-normal}.

Equation~\eqref{eq-normal} is a force balance relation ensuring that, when evaluated over the mapped ensemble, the AA model and the approximate CG potential generate the same mean force along each force field basis vector.\cite{Mullinax:2009b}
Consequently, Eq.~\eqref{eq-normal} may be interpreted as a generalization of the Yvon-Born-Green equation, which allows BOCS to perform variational force-matching without force information.\cite{Rudzinski:2015bb}
However, while this capability has been implemented in BOCS for bonded, pair, and LD potentials, it has not yet been implemented for SG potentials.

\section{Computational Methods}
\subsection{Modifications to BOCS version 5.0}
\subsubsection{Computational workflow}
\begin{figure}[h!]
\ifShowFig
	\includegraphics[width=3.25 in]{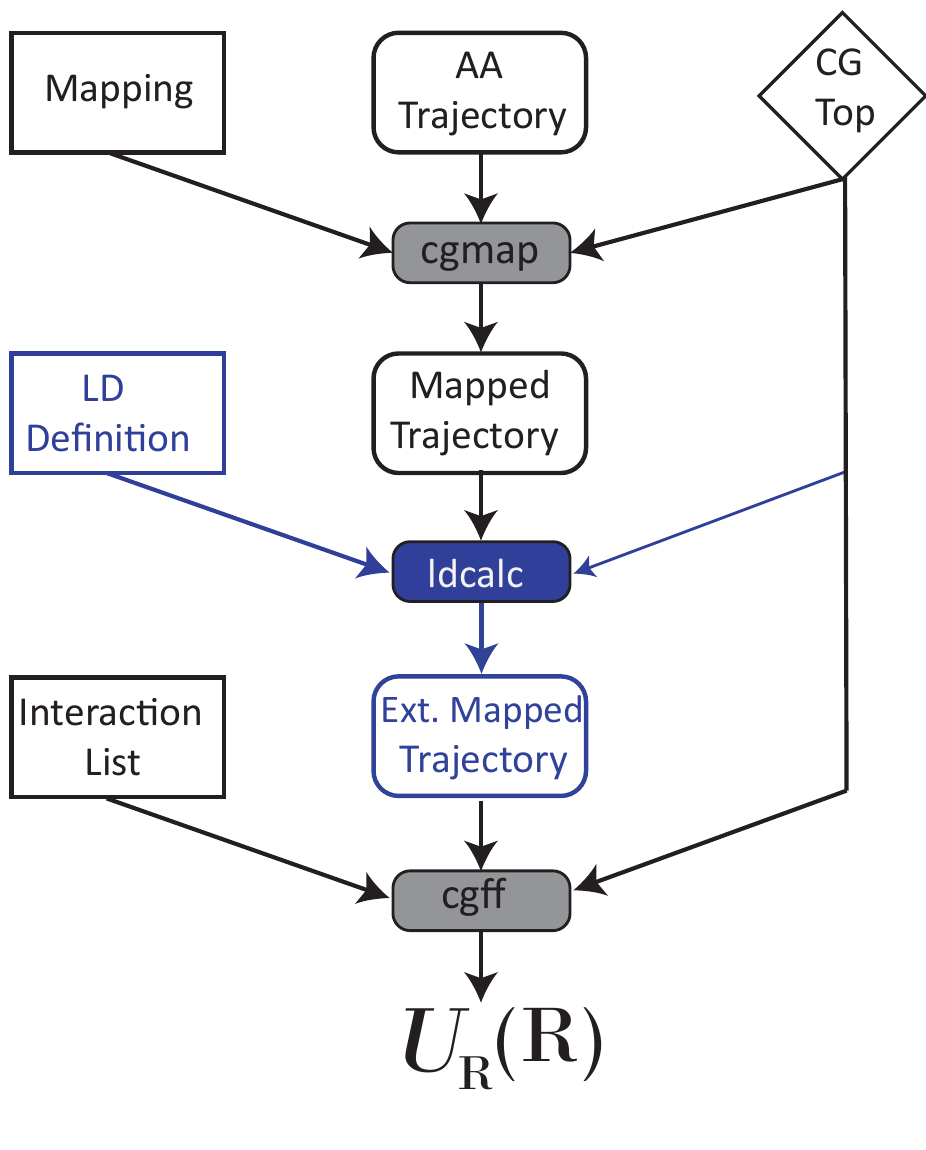}
\fi
	\caption{
	Basic workflow for parameterizing the CG interaction potential, $\UR(\bR)$, with BOCS. 
	Shaded boxes with rounded corners indicate BOCS programs.
	Unshaded boxes with rounded corners indicate the input AA trajectory and how it is processed through the workflow. 
	Rectangles with sharp corners indicate user provided input files, while the diamond indicates the CG topology.
	We highlight in blue the new BOCS components that are required for parameterizing LD or SG potentials.
	}
	\label{fig:BOCS_workflow}
\end{figure}

BOCS implements bottom-up approaches for parameterizing the approximate interaction potential, $U(\bR,V)$, in Eq.~\eqref{def-U}.
We have previously described the software for parameterizing the volume potential, $\UV(V)$, via self-consistent pressure-matching.\cite{Dunn:2015wn} 
Consequently, we only describe the software for parameterizing the approximate interaction potential, $\UR(\bR) \equiv \Ub(\bR) + \Unb(\bR)$.

\newcommand{\rVI}{{\rm V}_I}
Figure~\ref{fig:BOCS_workflow} illustrates the basic workflow for parameterizing $\UR(\bR)$.
This workflow begins with an AA trajectory that must include the atomic configuration, $\br_t = \{\bri(t)\}$, at one or more times, $t$.
The AA trajectory may also include the net force, $\bfi(\br_t)$, on each atom, $i$.
The \verb|cgmap| tool reads this AA trajectory file, as well as files that define the CG mapping, $\bM$, and the topology of the CG model.
The mapping file identifies the atoms that are associated with each site and also the corresponding mapping coefficients, $\cIi$.
The \verb|cgmap| tool is currently compatible with non-overlapping maps that associate the CG sites, $I = 1, \ldots, N$, with disjoint sets, $\{\rVI\}_{I=1, \ldots, N}$, of 1 or more atoms, i.e., each atom can contribute to 0 or 1 sites. 
Conversely, the CG topology file specifies the number and types of the sites and molecules in the CG model.
This topology file also specifies the composition and bonded connectivity of each molecule type, as well as the excluded intramolecular interactions.  
The \verb|cgmap| tool then outputs a CG trajectory file with the mapped representation, $\bR_t = \bM(\br_t)$, of each AA configuration, $\br_t$.
If the AA trajectory file includes force information, then the mapped trajectory file will also include the mapped force, $\bfI(\br_t)$, on each CG site, $I$.
For the non-overlapping maps that are currently supported, $\bfI(\br_t) = \sum_{i\in\rVI} \bfi(\br_t)$, is simply the net force on the atomic group, $\rVI$, associated with site $I$.\cite{Noid:2008a}
If  the approximate CG potential, $\UR$, does not include LD or SG terms, then \verb|cgff| can use this mapped trajectory to determine $\UR$.

If $\UR$ does include LD or SG terms, then we employ the \verb|ldcalc| tool to supplement the mapped trajectory with the necessary local density information.
In addition to the mapped trajectory and the CG topology, \verb|ldcalc| also reads a file that specifies the weighting function, $\wbar_{\tau|\tau'}$, and length-scale, $r_{\tau|\tau';\rm c}$, for the local density associated with each ordered pair, $\{\tau|\tau'\}$, of site types in the CG model. 
The \verb|ldcalc| tool then calculates the local density, $\rho_{\tau|I}(\bR_t)$, and the gradient, $\dI\rho_{\tau|I}(\bR_t)$, for each site type, $\tau$, around each CG site, $I$, in each mapped configuration, $\bR_t$.
The \verb|ldcalc| tool outputs an extended mapped trajectory that includes this additional information.

Finally, the \verb|cgff| tool determines the coefficients, $\phi_*$, for the interaction potential, $\UR(\phi_*)$, that optimally approximates the configuration-dependence of the exact potential, $W$.
In addition to the (original or extended) mapped trajectory and the CG topology file, the \verb|cgff| tool reads a file that specifies the various terms in $\UR$, as well as the options and hyperparameters employed in the variational force-matching calculation. 
The \verb|cgff| tool calculates the correlation matrix, $\GDDp$, and the force vector, $\bD$, that are defined by Eqs.~\eqref{def-GDDp} and \eqref{def-bD}. 
The \verb|cgff| tool then solves the normal equations, Eq.~\eqref{eq-normal} for the optimal coefficients, $\phi_*$. 
BOCS provides additional Python scripts and utilities that format $\UR$ for simulations in GROMACS or LAMMPS.
However, recent versions of GROMACS have not supported non-bonded potentials defined by flexible spline basis functions. 
Moreover, CG potentials with LD or SG terms must be simulated with LAMMPS.

We have worked with LAMMPS developers to extend PKG-BOCS in the mainline LAMMPS distribution for simulating LD and SG potentials.
PKG-BOCS can also be used as an analysis tool for calculating the local densities and their gradients around each particle in simulations that do not employ LD or SG potentials. 
The key new component in PKG-BOCS is a pair\_style (ldd) that enables users to combine LD and SG potentials with other pair\_styles. 
Our original implementation, which was used for the present calculations and will be hosted on our GitHub page,\cite{GitHubLammpsBocs} also introduced a new ldd atom\_style.
However, the ldd atom\_style was refactored into the ldd pair\_style in the mainline LAMMPS distribution. 
We have contributed documentation, unit tests, and examples for PKG-BOCS with the LAMMPS distribution.
We are distributing documentation on the Noid group GitHub page for parameterizing LD and SG potentials with BOCS version 5.0,\cite{GitHubBocs} as well as tutorials for simulating LD potentials with PKG-BOCS.

\subsubsection{Software implementation and ecosystem}
The \verb|cgff| tool provides an efficient C implementation of force-matching that scales well to large systems and long trajectories. 
While our original implementation calculated the many-body contributions to $\GDDp$ via a time-consuming triple loop over particles, the current version of \verb|cgff| calculates $\GDDp$ much more efficiently by directly employing the force field basis vectors. 
Moreover, \verb|cgff| employs an ``embarrassingly parallel'' OpenMPI implementation that reduces the computational time for computing $\GDDp$ and $\bD$ by a factor of $P$ by distributing the sampled configurations over $P$ independent processors.

BOCS employs its own file- and data-structures for treating topology and trajectory information.
Because it was initially developed from GROMACS software,\cite{Lindahl:2001,vanderSpoel:2005,Abraham:2015wn} BOCS originally enjoyed native compatibility with GROMACS topology and trajectory files, e.g., binary .tpr and .trr files. 
Subsequently, we developed stand-alone file formats and datastructures that no longer rely upon GROMACS installations.
BOCS now treats CG topology information with a compact, human-readable .btp file format that is quite similar to a dumped .tpr file.
While it can directly interface with conventional trajectory files from GROMACS and LAMMPS, BOCS employs a text, human-readable .btj file to store the LD information in the extended mapped trajectory.
Additionally, BOCS can read LAMMPS trajectory files that have been supplemented with LD information.

We have developed a \verb|translator| tool for porting topology and trajectory information into the BOCS format.
The \verb|translator| tool can convert dumped .tpr files from GROMACS into .btp format. 
The \verb|translator| tool can also convert GROMACS and LAMMPS trajectory files into the .btj format, as well as interconvert between these trajectory formats.
This tool provides BOCS with considerable flexibility, while still enjoying the benefits of the  GROMACS datastructures.
We anticipate regularly updating the \verb|translator| tool to maintain compatibility with future versions of these packages.

Finally, BOCS also provides a variety of tools for analysis and for facilitating simulations. 
For instance, the \verb|LDPDF| tool computes and outputs the sampled distribution for each type of local density and its gradient.
Similarly, the \verb|cgff| tool employs the g-YBG framework to analyze the cross-correlations between different degrees of freedom and to quantify their contribution to the calculated potential.
We have also developed new Python scripts \verb|LD_table_prep_from_spline.py| and \verb|SG_table_prep_from_spline.py| that format the calculated LD and SG potentials for simulations in LAMMPS.

\FloatBarrier

\subsubsection{Weighting functions}
\label{SubsubSec-Methods-WeightingFunctions}
The local density, $\rhoI$, around each particle, $I$, is defined by a normalized weighting function, $\wbar(r)$, according to Eq.~\eqref{def-rhoI}. 
We have implemented 5 different weighting functions into BOCS and LAMMPS.
Each function is determined by a non-increasing function, $w(r)$, that decreases from 1 to 0 over the interval $0 \le r \le \rc$ and vanishes for $r \ge \rc$.
Each function is then normalized such that $\int_0^{\rc} 4\pi r^2 \dd r \wbar(r) = 1$. 
The SI provides explicit expressions for each function.

\begin{figure}[h!]
 \ifShowFig
	\includegraphics[width=3.25 in]{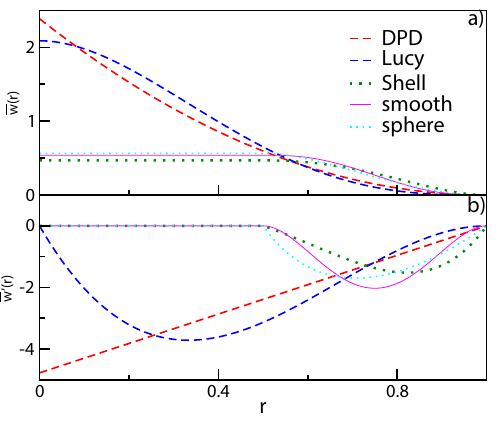}
\fi
	\caption{Normalized weighting functions implemented in BOCS and LAMMPS.
	We have plotted $\wbar(r)$ and $\wbar'(r)$ in dimensionless units by rescaling distances with respect to $\rc$, i.e., $r \to r/\rc$. 
	We have set $r_0 = \rc/2$ for the Shell, smooth, and sphere weighting functions. 
	}
	\label{fig:BOCS_w}
\end{figure}

Figure~\ref{fig:BOCS_w}a plots the 5 normalized weighting functions, $\wbar(r)$.
The dashed red curve plots the quadratic weighting function that is often employed in dissipative particle dynamics models.\cite{groot_dissipative_1997,Pagonabarraga:2001aa,Warren:2003}  
The dashed blue curve plots the Lucy weighting function\cite{Lucy:1977,monaghan_smoothed_2005} that has been used in several MS-CG models with LD and SG potentials.\cite{Moore:2016wn,Wagner:2017xx,Delyser:2019ld,Delyser:2020int,Delyser:2022wn} 
These weighting functions both monotonically decay over the interval $0 \le r \le \rc$.
The remaining weighting functions are all constant over an inner interval $0 \le r \le r_0$ and then decay from $r_0 \le r \le \rc$.
For concreteness, we have set $r_0 = \rc / 2$ in Fig.~\ref{fig:BOCS_w}.
The dotted green curve plots the weighting function employed by Sanyal and Shell.\cite{Sanyal:2016md,sanyal_transferable_2018}
The dotted cyan curve presents the `sphere' weighting function, which quantifies the fraction of overlap between two spheres that are separated by a distance $r$.
Finally, the thin pink curve presents a weighting function that decays more smoothly.

Figure~\ref{fig:BOCS_w}b plots the corresponding derivatives, $\wbar'(r)$, that directly contribute to the forces in Eq.~\eqref{def-bFIJnb}.
This derivative vanishes at $\rc$ for all weighting functions in order to ensure that the LD force is continuous. 
However, the higher order derivatives distinguish the various weighting functions.
The second derivative of the Lucy weighting function vanishes at $\rc$ and is continuous for all $r$.
In contrast, the second derivative of the DPD weighting function has a constant finite value over the interval $0 \le r < \rc$, but is discontinuous at $\rc$.
Similarly, the Shell and sphere weighting functions have discontinuous second derivatives at both $r_0$ and $\rc$.
The smooth weighting function has continuous second derivatives for all $r$ that vanish for $r \le r_0$ and $r \ge \rc$.

\subsection{Testing the new BOCS}
\label{SubSec-Methods-TestingBocs}
We report three tests that validate our implementation of LD and SG potentials in LAMMPS and BOCS.
In each test we employed PKG-BOCS to simulate a known interaction potential, $\UR(\bR)$, in LAMMPS.
We then employed BOCS to perform the variational force-matching calculation with the simulated configurations and net forces. 
All force-matching calculations employed BOCS version 5.0, which is available at the Noid group GitHub page.\cite{GitHubBocs}
All LAMMPS calculations employed the version that is hosted on the Noid group GitHub page at noid-group/lammps-BOCS2026.\cite{GitHubLammpsBocs}
Specifically, we created a branch, ``BOCS-2026LD'', within this repository to distribute the src code and documentation for the reported calculations. 
We have obtained equivalent results with the final version of PKG-BOCS that has been merged into the lammps/lammps:develop version, which is hosted on the LAMMPS GitHub page at https://github.com/lammps/lammps.

Each LAMMPS simulation employed a verlet neighbor list and three-dimensional periodic boundary conditions. 
We set the inner cut-off of the Shell, sphere, and smooth weighting functions to $r_0 = 0$ for simplicity.
Each force-matching calculation solved the corresponding normal equations via singular value decomposition with the hyperparameters listed in the SI.

\subsubsection{Test 1: 400 molecules with linear LD potential}
The first test considered $N = 400$ equivalent spherical particles that interacted via the potential 
$\UR(\bR) = \sum_I^N \Urho (\rhoI)$, 
where 
$\Urho(\rhoL) = -k \rhoL$ with $k = 1 $~kJ $\cdot$ mol$\inv~\cdot$ nm$^3$.
We defined the local density, $\rhoI$, around each molecule, $I$, with a weighting function, $\wbar(r)$, that vanished for $r \ge \rc = $ 1.0~nm.
We considered 5 different variations on this test that employed the 5 weighting functions, $\wbar(r)$, described in Section~\ref{SubsubSec-Methods-WeightingFunctions}.

\newcommand{\rlist}{r_{\rm list}}

We created the initial configuration by randomly placing the 400 particles within a sphere of radius 1.5~nm at the center of a cubic box with sides of length $L = $ 10~nm. 
We simulated this initial condition for 10 ns, while using a 1~fs time step to integrate the equations of motion and a neighbor list with 
$\rlist = 2.0$~nm. 
This simulation employed the default fix nvt settings to sample the canonical ensemble at $T = $ 270~K via a Nos\'e-Hoover chain thermostat\cite{Nose:1984, Hoover:1985,Martyna:1992wn} with a chain of length $n = $ 3 and relaxation time scale of 100~fs.
We sampled thermodynamic and trajectory information every 500 time steps.

We employed two different basis sets for the variational force-matching calculation. 
The first basis set employed only LD force functions, $\Frho(\rhoL)$.
In this case, we represented $\Frho(\rhoL)$ with 4\textsuperscript{th} order (cubic) splines, while employing a knot spacing and domain tailored to the local density range sampled by the LAMMPS simulation.
The second basis set employed only pair force functions, $F_2(r)$. 
We represented  $F_2(r)$ over the interval $0.001\text{~nm} \leq r \leq 1.5$~nm with 4\textsuperscript{th} order (cubic) splines, while employing a knot spacing $\de r=0.01$~nm.
The SI provides more details about these calculations.

\subsubsection{Test 2: 2 particles with LD or SG potentials}
\label{SubSubSec-Methods-Test2}
The second test considered $N = $ 2 equivalent spherical particles and two distinct cases.  
In the first LD case the particles interacted via the potential   
$
	\UR(\bR) 
		= 
			\Uld(\bR) 
			= 
				\sum_I^N U_{\rho}(\rho_I)
	,
$ 
where 
$
	\Urho(\rhoL) = A\rhoL^2 + B\rhoL + C
$
with 
$A =  2$~kcal $ \cdot~\text{\AA}^6 \cdot $ mol$^{-1}$, 
$B =  -4.4$~kcal $\cdot \text{~\AA}^3 \cdot $~ mol$^{-1}$, and 
$C =  2.42$~kcal $\cdot$ mol$^{-1}$. 
In the second SG case the particles interacted via the potential
	$
	\UR(\bR) = 
	\Usg(\bR) 
	= 
	\sum_I^N \Udel(\rhoI)|\dI \rhoI|^2
	,
	$ 
	where  
	$
	\Udel(\rhoL) = \tilde{A}\rhoL + \tilde{B}
	$
	with  
	$\tilde{A} =  32$~kcal $\cdot~\text{\AA}^{11} \cdot$ \pmol~and 
	$\tilde{B} =  -64$~kcal $\cdot~\text{\AA}^8 \cdot$ \pmol.
We considered 5 variants of each case that defined the local density according to the 5   weighting functions, $\wbar(r)$, described in Section~\ref{SubsubSec-Methods-WeightingFunctions}.
The LD and SG cases defined the local density length scale by $\rc = $ 1.2~\AA~and 2.0~\AA, respectively.

In each case we initially placed the two particles 1~\AA~away from each other.
We then simulated them in a cubic box with sides of fixed length $L=7$~\AA~for a total of 6.3 $\times 10^9$ time steps, while integrating the equations of motion with a 1 fs timestep.
These simulations employed a Langevin\cite{dunweg1991brownian, brunger1984stochastic} thermostat with a 50 fs damping factor to sample the canonical ensemble at $T = $ 270~K.
The simulations for the LD and SG cases employed neighbor lists with $\rlist = 2.4$~\AA \, and 4~\AA, respectively. 
In both cases, we checked at every time step whether the neighbor list needed to be rebuilt.
We sampled every 500$^{\rm th}$ configuration for analysis.
We determined the conservative forces by postprocessing these sampled configurations with the LAMMPS fix nvt command.

In both LD and SG cases, we performed two independent force-matching calculations with different basis sets:
\begin{enumerate}
\item
The first force-matching calculation employed basis functions of the local density, $\rhoL$.
In the LD case we represented the LD force function, $\Frho(\rhoL)$, with  4\textsuperscript{th} order (cubic) spline functions.
These calculations employed approximately 10 evenly spaced knots over the range of simulated local densities.
In the SG case we represented the SG coefficient function, $\Udel(\rhoL)$, with a single 2\textsuperscript{nd} order (linear) spline function. 
We quantified the error in the force-matching calculations for the LD case by directly comparing the simulated LD force function with the calculated LD force function.
We quantified the error in the force-matching calculations for the SG case by  comparing the net forces that are generated by the simulated and calculated SG potentials. 
We computed these net forces by post-processing the sampled configurations with the corresponding SG potentials.
\item
The second force-matching calculation employed basis functions of the pair distance. 
In the LD case, we represented the  pair force function, $\Ftwo(r)$, by linear spline functions with knots spaced every $\de r = $ 0.0001~nm over the interval $0.0 \text{~nm}\leq r \leq 0.2$~nm.
In the SG case, we represented  $\Ftwo(r)$ by cubic spline functions with knots spaced every $\de r = $ 0.0005~nm over the interval $0.0 \text{~nm}\leq r \leq 0.2$~nm.
The only exception occurred for the Lucy function.
In this case we represented  $\Ftwo(r)$ by cubic spline functions with knots spaced every $\de r = $ 0.002~nm over the interval $0.0 \text{~nm}\leq r \leq 0.2$~nm. 
We quantified the error in these force-matching calculations by averaging the difference between the standard function and the calculated function within each knot interval.
Specifically, we computed this difference at 300~evenly spaced points within each knot interval and divided the sum of these differences by 300.
\end{enumerate}

The simulated ensembles for this second test also provided an independent validation of the LD and SG potentials implemented in LAMMPS.
For the present case of $N = $ 2 particles, the total potential may be re-expressed in terms of an effective pair potential, $U(\bR) = \tUtwo(R_{12})$, where $R_{12}$ is the distance between the two particles. 
Consequently, the canonical probability (density) for finding the two particles separated by a distance $r$ is $P_{\rm r}(r)  \propto r^2 \exp[-\bt \tUtwo(r)]$. 
In particular, the regime $r > \rc$ corresponds to free particles for which $P_{\rm r}(r)  \propto r^2$. 
In the present work we determined the constant of proportionality, $C$, by fitting the sampled distribution of pair distances to $P_{\rm r}(r)  = C r^2 \exp[-\bt \tUtwo(r)]$.
This canonical pair distribution then determines the canonical distribution of local densities, $P_{\rho}(\rhoL)$, which may be decomposed :
\begin{equation}
P_{\rho}(\rhoL) = \Pb(\rhoL) + \Pf \de(\rhoL)
\end{equation}
where $\Pb$ and $\Pf = 1 - \Pb$ correspond to contributions from bound and free configurations, respectively. 
Additionally, in the present case of $N = $ 2 particles, the local density is simply $\rhoL = \rho_1 = \rho_2 = \overline{w}(R_{12})$.
Because $\wbar$ monotonically decreases over the interval $0 \le r \le \rc$, there exists a one-to-one relationship between the interparticle distance, $0 \le R_{12} \le \rc$, and the local density, $0 \le \rhoL \le \rhomax$.
This allows us to determine the bound contribution to the canonical probability distribution:
\begin{equation}
\Pb(\rhoL) = P_{\rm r}(r) \left. \left|\frac{\dd\wbar}{\dd r}\right|\inv \right|_{r = R_{\rhoL}},
\end{equation}
where $R_{\rhoL}$ is the distance corresponding to $\rhoL$, i.e., $\rhoL = \wbar(R_{\rhoL})$.

\subsubsection{Test 3: 125 molecules with LD and SG potentials}
\label{subsubsec-Methods-Test3}
The third test considered a system with 125 molecules and 4 different types of CG sites, $\tau = $  A, B, C, or D.
Each molecule consisted of 9 sites with a branched topology.
The interaction potential included both bonded and nonbonded contributions, $\UR(\bR) = \Ub(\bR) + \Unb(\bR)$.
The bonded potential, $\Ub$, treated 11 distinct types of interactions, including 4 bond potentials, 5 angle potentials, and 2 dihedral potentials.
The nonbonded potential, $\Unb$, treated 42 distinct types of interactions, including 10 pair potentials, 16 LD potentials, and 16 SG potentials. 
The nonbonded potential governed both intermolecular interactions and also intramolecular interactions between sites that were separated by 4 or more bonds, i.e., we excluded nonbonded interactions between intramolecular 1-2, 1-3, and 1-4 pairs.  
We represented all 42 inter-molecular interactions with cubic spline functions on a uniformly spaced grid.
We defined these potential functions by smoothing and scaling potential functions that were previously calculated for a 1-site CG model of acetic acid.\cite{Delyser:2022wn}
We define the model potential more explicitly in Section~\ref{subsubsec-Results-Test3}. 

In order to simplify the presentation, we did not employ distinct functions for all 53  interaction potentials. 
In particular, we defined the 4 distinct bond potentials and the 5 distinct angle potentials with the same harmonic functions of the bond and angle, respectively.
Additionally, we defined all 10 distinct pair potentials with the same function.

We adopted similar simplifications for defining the LD and SG potentials. 
While this model treats 16 distinct types of local density, we adopted the same Lucy weighting function with $\rc = $ 0.65~nm to define the local density, $\rho_{\tau|I}$, of $\tau$ type sites around each site, $I$. 
The LD potentials, $U_{\rho \tau|\tau_I}$, for site $I$ depend upon the type, $\tau_I$, of site $I$ but not the type, $\tau$, of the surrounding particles. 
Similarly, the SG coefficients, $U_{\del \tau|\tau_I}$, for site $I$ depend upon $\tau_I$ but not $\tau$.
Furthermore, the LD potential and SG coefficient functions for different site types are related by scalar multiples. 

An important technical detail is that we defined the LD and SG potentials such that their support was within the range of local densities sampled by the LAMMPS simulation.
If the range of sampled local densities does not completely cover the support of the LD and SG potentials, then the boundary conditions of the spline basis functions employed in the variational force-matching calculation are not consistent with the simulated LD and SG potentials at the boundary of the sampled local densities, i.e., the basis functions vanish at this boundary but the simulated potentials do not.
This not only results in slight errors at the boundaries of the calculated LD and SG functions, but also generates oscillations in the calculated pair forces at $\rc$.
Essentially the discrepancy at the boundaries of the LD and SG functions couple to the boundary of the LD contribution to the pair force.
Nevertheless, the calculated total potential accurately matches the simulated net forces, i.e., $\chi^2 \approx 0.$
Consequently, we suspect that these considerations are not important in the case of systematic coarse-graining because they likely introduce little error in reproducing the true mean forces.

We generated the initial configuration for this third test by employing the LAMMPS \verb|create_atoms| command with a template to randomly place the molecules in a cubic box with sides of length $L = $ 
4~nm.
After performing a brief energy minimization, we simulated the system for 
22~ns with the constant nvt fix, while employing a 
~1 fs time step to integrate the equations of motion for a Nos\'e-Hoover chain\cite{Nose:1984, Hoover:1985, Martyna:1992wn} with the default chain length, $n = $ 3, and a damping time of 100~fs.
The simulations employed a 3.0~nm neighbor list, which was checked at every time step.
We sampled every 500$^{\rm th}$ configuration from this simulation.

The variational force-matching calculation simultaneously determined 25,109 potential parameters. 
This calculation employed the same 5,957 spline basis functions that were used to define the non-bonded potentials.
This calculation also employed a total of 19,152  spline basis functions to represent the bonded potentials. 
We tailored the knot spacings and polynomial degrees to accurately describe each bonded potential.
The net forces do not determine unique LD potentials.
Rather the non-bonded potentials obtained from force-matching are members of a family of equivalent potentials that generate identical net forces.\cite{Delyser:2019ld}
In the following, we report the member of this family that most nearly matches the simulated potentials.
We identify this member based upon an automated procedure previously described in Refs.~\citenum{Szukalo:2023} and \citenum{Lesniewski:2026a}.

\subsection{Applying the new BOCS}
\label{SubSec-Methods-ApplyingBocs}
We illustrate this new software by developing and assessing 1-site CG water models based upon the AA SPC/E model.\cite{Berendsen:1987}
We refer readers to our recent study\cite{Lesniewski:2026a} for a much more detailed analysis of CG water models based upon the 4-site TIP4P model.\cite{jorgensen_comparison_1983} 
As described above in Sec.~\ref{SubSec-Methods-TestingBocs}, we performed all force-matching calculations with BOCS version 5.0, which is available at the Noid group GitHub page.\cite{GitHubBocs}
Similarly, we performed all LAMMPS simulations with the ``BOCS-2026LD'' version that is available on the Noid group GitHub page.\cite{GitHubLammpsBocs}
We have obtained equivalent results with the developmental version of LAMMPS that is hosted at https://github.com/lammps/lammps.	

\subsubsection{SPC/E model}

We performed two simulations of $N = $ 5000 water molecules with GROMACS/2019.6, while propagating dynamics with the md leapfrog integrator, a 2 fs time step, and periodic boundary conditions in all 3 directions.\cite{Lindahl:2001,vanderSpoel:2005,Abraham:2015wn} 
We described intermolecular interactions with the SPC/E water model,\cite{Berendsen:1987} while employing the LINCS\cite{Hess:1997} algorithm to rigidly constrain the geometry of each molecule.
We treated long-ranged electrostatic interactions with the particle mesh Ewald (PME) method,\cite{Darden:1993} while truncating dispersive interactions and short-range PME contributions at 1.55~nm. 
We did not treat dispersion corrections in order to ensure consistency between bulk and slab simulations.
We sampled energy fluctuations with the Bussi thermostat,\cite{Bussi:2007} while employing a 0.5 ps relaxation time. 
We sampled volume fluctuations with the Parinello-Rahman\cite{Parrinello:1980} barostat, while employing a relaxation time of 5 ps and a compressibility of 4.5 $\times10^{-5}$ bar$^{-1}$.

We first simulated bulk water in the constant NPT ensemble under ambient conditions, $\Tref = $ 300~K and $\Pref = $ 1~bar.
We constructed an initial condition for this NPT simulation by employing \verb|gmx insert-molecules| to insert $N = $ 5000 water molecules into a cubic box with sides of length $L=5.32502~$nm and conducting a brief energy minimization.
We simulated this configuration for 34~ns and treated the first 4~ns as equilibration.
We extended the z dimension of the last configuration from this constant NPT simulation to $z=16.0$~nm.
We then simulated the resulting configuration in the constant NVT ensemble at $\Tref = $ 300~K for 34~ns and treated the first 4~ns as equilibration.
We sampled each production simulation every 500 timesteps, corresponding to a rate of 1 frame/ps.

\subsubsection{CG models}	
We parameterized three different 1-site CG water models from these SPC/E simulations via variational force-matching, while employing the default settings to solve Eq.~\eqref{eq-normal}.

The first two CG models both employed the same pair-additive form for the approximate CG potential, $U(\bR) = \sum_{(I,J)} U_2(\RIJ).$
We represented the pair force function, $F_2(r)$, in each case with 4$^{\rm th}$ order (cubic) splines, while employing a knot spacing of $\delta r = 0.02$~nm over the interval 0.0~nm $\leq r \leq$ 1.4~nm.
We parameterized the PAIR-BULK and PAIR-SLAB models based upon the bulk and slab SPC/E simulations, respectively.

The third CG model included both pair and LD terms in the  approximate potential, $U(\bR) = \sum_{(I,J)} U_2(\RIJ) + \sumIN \Urho(\rhoI)$.
We parameterized this LD-BULK model from the bulk SPC/E simulation.
We represented the pair force function, $F_2(r)$, with 4$^{\rm th}$ order (cubic) splines, while employing a knot spacing of $\delta r = 0.01$~nm over the interval 0.0~nm $\leq r \leq$ 1.4~nm.
We defined the local density, $\rhoI$, around each site, $I$, with a normalized Lucy function, while including the self-term in Eq.~\eqref{def-rhoI}. 
We employed an automated search\cite{Lesniewski:2026a} to identify the LD length-scale, $\rc = $ 0.392~nm, that best reproduced the bulk density of the SPC/E model. 
We represented the LD force function, $\Frho(\rhoL)$, with 4\tsup{th} order (cubic) splines, while employing a knot-spacing $\delta \rhoL = $ 0.1~nm\tsup{-3} over the range of local densities sampled by the bulk SPC/E simulation. 
We extrapolated $\Frho(\rhoL)$ outside of this range according to the automated procedure previously described in Ref.~\citenum{Lesniewski:2026a}.

We simulated these CG models with LAMMPS, while propagating dynamics with a 2~fs timestep and employing periodic boundary conditions in all 3 directions.
Constant NVT simulations employed a Nos\'e-Hoover chain thermostat,\cite{Nose:1984, Hoover:1985,Martyna:1992wn} while employing a 200~fs relaxation time and the default chain length $n = $ 3.
Constant NPT simulations employed the equations of motion developed by Shinoda et al.\cite{shinoda_rapid_2004, Martyna:1994nd, parrinello_polymorphic_1981}
All interactions were truncated at the cutoff $\rcut =1.4$~nm.
We employed a neighborlist with a cutoff of $r_{\rm list} = 3 $~nm, which was checked at every time step. 
We obtained the initial configurations for these simulations by mapping the last configuration of the corresponding SPC/E simulation to the CG resolution.

\FloatBarrier
\section{Results and Discussion}

\subsection{Benchmark tests}
\newcommand{\URFM}{\UR^{\rm FM}}
\newcommand{\Ufm}{U_{\rm FM}}
\newcommand{\Usim}{U_{\rm sim}}
In this subsection we present three benchmark tests that illustrate and validate our implementation of LD and SG potentials in LAMMPS and BOCS.
In each case, we first simulated a known potential, $\Usim$, in LAMMPS with the PKG-BOCS package.
We then employed BOCS to calculate a potential, $\Ufm$, for the same resolution via variational force-matching with the simulated configurations and forces. 
Because we are not coarse-graining the underlying model, the known simulated potential minimizes the force-matching functional, i.e., $\chi^2[\Usim] = 0$.
Consequently, the difference, $\Usim(\bR) - \Ufm(\bR)$, between the simulated and calculated potentials directly assesses the software that we implemented in LAMMPS and BOCS. 
In addition to demonstrating the accuracy of our implementation, these studies provide insight into the forces generated by LD and SG potentials.
In particular, we highlight the impact of the LD weighting function, $\wbar$, upon these forces.

\subsubsection{Test 1: 400 particles with linear LD potential}
\label{subsubsec-Results-Test1}

\renewcommand{\bar}[1]{\overline{#1}}
\renewcommand{\wbar}{\bar{w}(r)}
\newcommand{\wbarIJ}{\bar{w}(R_{IJ})}
\renewcommand{\wp}{\bar{w}'(r)}

\begin{figure}[h!]
\ifShowFig
	\includegraphics[width=3.25in]{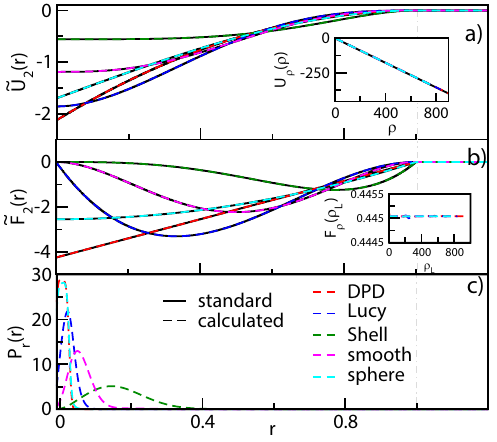}
\fi
	\caption{
	Analysis of $N = $ 400 molecules interacting via only LD potentials.
	The insets of panels a and b report the simulated LD potential, $\Urho$, and force function, $\Frho$. 
	The remainder of panels a and b report the corresponding effective pair potential, $\tUtwo$, and pair force, $\tFtwo$.
	Panel c reports the simulated pair distributions.
	The red, blue, green, pink, and cyan curves correspond to the DPD, Lucy, Shell, smooth, and sphere weighting functions, respectively.
	In panels a and b, the solid black curves present the standard functions that are defined by the simulated potential, $U$,
	while the dashed colored curves present the potentials and force functions determined via variational force-matching.
	We have scaled energies with respect to $\kT$ and distances with respect to $\rc$.
	}
	\label{fig:ToyModel1}
\end{figure}

We first consider a system of $N = $ 400 spherical particles that interact only via LD potentials:
\begin{equation}
\label{def-U1}
U(\bR) = \sumIN \Urho(\rhoI)		.
\end{equation}
We adopt the simplest possible LD potential,
\begin{equation}
\label{def-Urho1}
\Urho(\rhoL) = - k \rhoL	, 
\end{equation}
with $k = $ 0.44504,
which is plotted in the inset of Fig.~\ref{fig:ToyModel1}a.
Here we are employing reduced units that correspond to scaling energies by thermal energy, $\kT$, and scaling distances with respect to the LD length-scale, $\rc$.
The inset of Fig.~\ref{fig:ToyModel1}b presents the corresponding LD force function, $\Frho(\rhoL) = - \dd \Urho(\rhoL)/ \dd \rhoL = + k$.
The LD force function is independent of $\rhoL$ and always drives particles towards regions of higher local density.
Because Eq.~\eqref{def-U1} does not account for excluded volume, this interaction potential cannot realistically model molecular systems. 
Nevertheless, the LD potential defined by Eq.~\eqref{def-Urho1} may be useful for reducing the internal pressure of a more realistic  potential that already treated excluded volume.
Moreover, the simplicity of the present model provides transparent insight into the forces generated by LD potentials.

The definition of the local density in Eq.~\eqref{def-rhoI} implies that, if $\Urho(\rhoL)$ is an $m$\tsup{th} degree polynomial in $\rhoL$, then $\Urho$ can be represented as a sum of $m+1$-body terms.\cite{Warren:2003} 
Consequently, the total potential, Eq.~\eqref{def-U1}, can be re-expressed in terms of central pair potentials:
\begin{equation}
\label{eq-U1}
U(\bR) = \sum_{(I,J)} \tUtwo(\RIJ)		,
\end{equation}
where 
\begin{equation}
\label{def-tUtwo1}
\tUtwo(r) = -2 k \wbar .
\end{equation}
Figures~\ref{fig:ToyModel1}a and \ref{fig:ToyModel1}b present the corresponding pair potentials, $\tUtwo(r)$, and pair forces, $\tFtwo(r) = - \dd \tUtwo(r) / \dd r$, for the 5 weighting functions that we have implemented in BOCS and LAMMPS.
(We have set the inner radius of the Shell, smooth, and sphere weighting functions to $r_0 = $ 0 for simplicity.)
These curves emphasize that the interaction potential depends not only upon the LD potential, $\Urho$, but also the weighting function used to define the local density.
Given the fixed LD potential function, the DPD and Lucy weighting functions result in the most attractive pair potentials, while the Shell weighting function results in the least attractive potential.
The pair forces for the Shell, smooth, and Lucy weighting functions are most attractive at $r \approx $ 0.80,  0.55, and 0.37, respectively, and all vanish at $r = $ 0.
Conversely, the pair forces for the DPD and sphere weighting functions become increasingly attractive as $r$ decreases and exert the largest forces at $r = $ 0.

We employed LAMMPS to simulate the system of $N = $ 400 particles interacting with the LD potential $\Urho(\rhoL) = - k \rhoL$, while employing each of the five weighting functions described in Section~\ref{SubsubSec-Methods-WeightingFunctions}.
Figure~\ref{fig:ToyModel1}c presents the sampled distribution, $P_{\rm r}(r)$, of pair distances for each of the five simulations.
The simulated statistics follow the expected trends.
In particular, the DPD and sphere weighting functions result in the most compact aggregates, while the Shell weighting function results in more extended aggregates.

We employed BOCS to determine the simulated force functions via variational force-matching.
We performed this force-matching calculation with two different basis sets.
We first employed a basis set corresponding to local density forces.
The inset of Fig.~\ref{fig:ToyModel1}b demonstrates that BOCS quantitatively recovered the same simulated LD  force function, $\Frho(\rhoL) = + k$, from each of the 5 simulations.
We then employed a basis set corresponding to simple pair potentials.
The main panel of Fig.~\ref{fig:ToyModel1}b demonstrates that, in this case, BOCS quantitatively recovered the corresponding pair force for each of the 5 weighting functions.

\subsubsection{Test 2: Two particles with LD or SG potentials}
\label{subsubsec-Results-Test2}
\begin{figure}[h!]
\ifShowFig
	\includegraphics[width=5in]{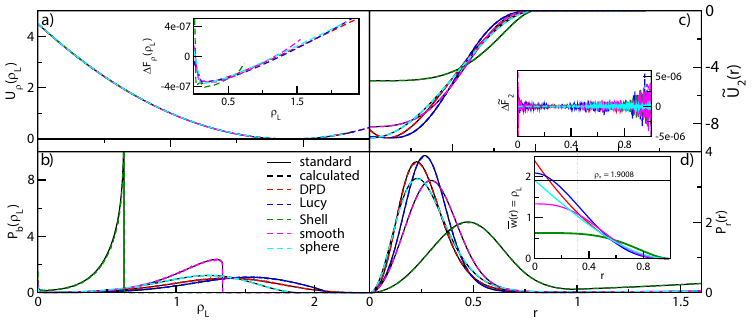}
\fi
	\caption{
	Analysis of $N = $ 2 particle simulations employing the LD potential, $\Urho(\rhoL)$, in Eq.~\eqref{def-Urho2}.
	Panel a presents $\Urho(\rhoL)$, while panel b presents the bound contribution, $\Pb(\rhoL)$, to the LD probability distribution.
	Panel c presents the corresponding effective pair potential, $\tUtwo(\RIJ)$, given by Eq.~\eqref{eq-tUtwo2}, while panel d presents the pair probability distribution, $P_{\rm r}(r)$.
	In panels a and c the solid black curves present the standard LD and pair potentials that are defined by the simulated potential, while the dashed curves present the corresponding potentials calculated via variational force-matching.
	The insets in these panels report the errors in the force-matching calculations.
	In panels b and d the solid black curves present the analytically determined, canonical  probability distributions, while the dashed colored curves present the simulated distributions.
	The inset in panel d presents the 5 local density weighting functions, $\wbar$, while the horizontal line indicates the equilibrium local density, $\rho_*$.
	The red, blue, green, pink, and cyan curves correspond to the DPD, Lucy, Shell, smooth, and sphere weighting functions, respectively.
	We have scaled energies with respect to $\kT$ and distances with respect to $\rc$.
		} 
	\label{fig:2PartLDonly}
\end{figure}

\paragraph{LD potential}
\label{para-LDpotl}
We next consider a system of $N = $ 2 spherical particles that interact only via  LD potentials according to Eq.~\eqref{def-U1}.
In this second test we consider the slightly more complex quadratic LD potential, 
\begin{equation}
\label{def-Urho2}
\Urho(\rhoL) = \half k \left( \rhoL - \rho_* \right)^2, 
\end{equation}
which is plotted in Fig.~\ref{fig:2PartLDonly}a.
The corresponding LD force function, 
	$
\Frho(\rhoL) 
	= 
		- \dd \Urho(\rhoL) / \dd \rhoL 
	= 
		- k \left( \rhoL - \rho_* \right)
	$  
generates attractive forces when $\rhoL < \rho_*$ and repulsive forces when $\rhoL > \rho_*$.
We have again scaled energies with respect to $\kT$ and distances with respect to $\rc$.

In the present case of $N = $ 2 particles, both particles have the same local density, $\rho_1 = \rho_2 = \rhoL$, which is determined by the weighting function, $\rhoL = \wb(R_{12})$.
The inset of Fig.~\ref{fig:2PartLDonly}d presents this relationship for the 5 weighting functions implemented in BOCS and LAMMPS.
(We have again set $r_0 = 0 $ for the Shell, smooth, and sphere weighting functions.)
When the particles are separated by $R_{12} > 1$, the local density vanishes, $\rhoL = \wb(R_{12}) = 0$.
Conversely, over the distance range $0 \le R_{12} \le 1$, $\rhoL$  increases as the particles approach and attains a maximum, $\rhomax$, at $R_{12} = 0$.  
Because the weighting functions have different shapes and, thus, different normalizations, the maximum local density differs among the weighting functions.
In the case of the DPD and Lucy weighting functions, the local density can exceed $\rho_*$ when particles are sufficiently close, i.e., $\rhoL > \rho_*$ for $R_{12}< r_*$.
In contrast, the local density is always less than $\rho_*$ for the Shell, smooth, and sphere weighting functions.

In general, a quadratic LD potential will generate 3-body forces. 
However, for the present case of $N = $ 2 particles, the total potential, $U$, can be expressed in terms of a single effective pair potential, $\tUtwo$, that explicitly reflects the weighting function: 
\begin{equation}
\label{eq-tUtwo2}
\tUtwo(\RIJ) = k \left[ \wbarIJ - \rho_* \right]^2	 . 	
\end{equation}
Figure~\ref{fig:2PartLDonly}c presents these pair potentials for the 5 weighting functions.
These pair potentials appear quite similar to the corresponding potentials in Fig.~\ref{fig:ToyModel1}a.
The greatest discrepancies arise for the DPD and Lucy weighting function.
In these cases, the pair potential generates repulsive forces at sufficiently small distances, $R_{12} < r_*$, for which $\rhoL > \rho_*$.
In contrast, the pair potentials is never repulsive for the Shell, smooth, and sphere weighting functions.

It is straightforward to analytically characterize the equilibrium Boltzmann distribution for $N = $ 2 particles.
Figure~\ref{fig:2PartLDonly}d plots the canonical probability density, $P_{\rm r}(r) \propto r^2 \exp[-\bt \tUtwo(r) ]$, for the two particles to be separated by a distance $r$.
The probability densities for the DPD and Lucy weighting functions are both sharply peaked near the distance $r_*$ corresponding to $\rho_*$.
Conversely, the Shell weighting function results in a much broader distribution that is peaked at considerably larger distances and, moreover, gives considerable weight to dissociated configurations with $r > 1$.
The relationship $\rhoL = \wbar$ then determines the corresponding local density distribution, $P_\rho(\rhoL) = \Pf \de(\rhoL) + \Pb(\rhoL)$, where $\Pf$ is the total probability for dissociated configurations, while $\Pb(\rhoL)$ is the equilibrium probability density for bound configurations.
Figure~\ref{fig:2PartLDonly}b presents $\Pb(\rhoL)$, which necessarily vanishes for $\rhoL > \overline w(0)$.
This distribution reflects not only the Boltzmann weight, $\exp[-\bt \Urho(\rhoL)]$, associated with the LD potential, but also the Jacobians associated with transforming from Cartesian coordinates to the pair distance, $R_{12}$, and then to the local density, $\rhoL$. 
Consequently, because it varies least rapidly with $r$, the Shell weighting function results in the broadest $r$ distribution and the sharpest $\rhoL$ distribution.

We employed LAMMPS to simulate Langevin dynamics for the LD potential in Eq.~\eqref{def-Urho2} with each of the 5 weighting functions.
The dashed curves in Figs.~\ref{fig:2PartLDonly}b and \ref{fig:2PartLDonly}d report the sampled distance and LD distributions, respectively.
The quantitative agreement between the analytically determined canonical distributions and the simulated distributions demonstrate the accuracy of our implementation of LD potentials in LAMMPS.

We then employed BOCS to determine the simulated force functions from the sampled forces. 
As in the preceding example, we performed two independent variational FM calculations.
The first FM calculation represented the forces with a LD force function, while the second represented the forces with a central pair force function.
The insets of Figs.~\ref{fig:2PartLDonly}a and \ref{fig:2PartLDonly}c report the errors in the calculated LD and pair force functions.
BOCS again determines these force functions very accurately.
In particular, the errors in the calculated pair force are very small and fluctuate around 0.
The errors in the calculated LD force are also very small and reflect the boundary conditions of the spline functions employed in the variational FM calculations.
Specifically, while the calculated LD force function necessarily vanishes for the largest and smallest local densities, Fig.~\ref{fig:2PartLDonly}a indicates that the simulated LD force function does not.

\paragraph{SG potential}
\label{para-SGpotl}
We now consider a system of $N = $ 2  spherical particles that interact only via SG potentials:
\begin{equation}
\label{def-U2b}
U(\bR) 
	\equiv 
		\sumIN \Udel(\rhoI) \left| \dI \rhoI \right|^2	
	= 
		2 \Udel(\rhoL) \left| \del \rhoL \right|^2	.
\end{equation}
The second expression simplifies $U$ for the case of $N = $ 2:  $\rhoL \equiv \rho_1 = \rho_2$ and $\del \rhoL \equiv \del_1 \rho_1 = - \del_2 \rho_2$.
Classical density functional theory and phenomenological mean field theories often describe the free energy cost of density gradients with similar functional forms where the coefficient of the square gradient is a positive constant, i.e., $\Udel(\rhoL) = \kappa > 0$.\cite{Evans:1979fk,Hansen:1990}
BOCS allows a more general treatment of the square gradient coefficient, $\Udel(\rhoL)$, that can be tuned to act specifically at interfaces.\cite{Delyser:2022wn}

\begin{figure}[h!]
\ifShowFig
	\includegraphics[width=5in]{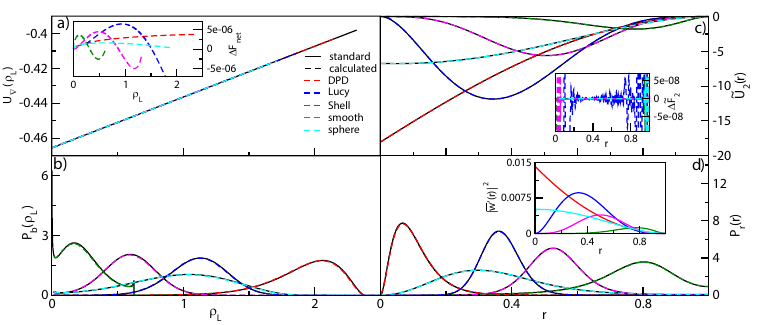}
\fi
	\caption{
		Analysis of $N = $ 2 particle simulations employing the SG potential defined by Eq.~\eqref{def-U2b}.
	Panel a presents the SG coefficient function, $\Udel(\rhoL)$, defined by Eq.~\eqref{def-Udel}, while panel b presents the bound contribution to the LD probability distribution, $\Pb(\rhoL)$.
	Panel c presents the corresponding effective pair potential, $\tUtwo(r)$, defined by Eq.~\eqref{eq-tUtwo-sg}, while panel d presents the pair probability distribution, $P_{\rm r}(r)$.
	In panels a and c the solid black curves present the standard SG coefficient and pair potentials that are defined by the simulated potentials, while the dashed curves present the corresponding functions calculated via variational force-matching.
	The insets in these panels present the errors in the force-matching calculations.
	In particular, the inset of panel a presents the error in the net force on each particle resulting from the force-matching calculation with the SG basis set, while the inset of panel b presents the error in the calculated pair force function.
	In panels b and d the solid black curves present the analytically determined, canonical probability distributions, while the dashed colored curves present the simulated distributions.
	The inset in panel d presents $\left| \bar{w}'(r) \right|^2$.
	The red, blue, green, pink, and cyan curves correspond to the DPD, Lucy, Shell, smooth, and sphere weighting functions, respectively.
	We have scaled energies with respect to $\kT$ and distances with respect to $\rc$.
		}
			\label{fig:2PartSGonly}
\end{figure}

We consider a linear SG coefficient function for this illustration,
\begin{equation}
\label{def-Udel}
\Udel(\rhoL) \equiv k \left( \rhoL - \rho_* \right) 	,
\end{equation}
which is plotted in Fig.~\ref{fig:2PartSGonly}a.
Note that $\Udel$ is not a potential and has dimensions of $\kT / \rc^8$.
(We have again scaled energies and distances with respect to $\kT$ and $\rc$, respectively.)
Because $\Udel(\rhoL) < 0$ over the relevant density range, $U$ stabilizes density gradients.

\newcommand{\Rtwo}{R_{12}}
For the present case of $N = $ 2 particles, $\left| \del \rhoL \right|^2 = \left| \bar{w}'(\Rtwo) \right|^2$.
While the inset of Fig.~\ref{fig:2PartLDonly}d plots $\rhoL = \bar{w}(\Rtwo)$ for the 5 weighting functions implemented in BOCS, the inset of Fig.~\ref{fig:2PartSGonly}d plots the corresponding square gradients, $\left| \bar{w}'(\Rtwo) \right|^2$, as a function of the pair distance, $R_{12}=r$.
In the case of the DPD and sphere weighting functions, $\left| \bar{w}'(r) \right|^2$ monotonically increases as $r$ decreases and approaches its maximum as $r \to 0$.
However, $\left| \bar{w}'(r) \right|^2$ is significantly smaller and varies less rapidly with distance for the sphere weighting function. 
Conversely, $\left| \bar{w}'(r) \right|^2$ attains a maximum at a finite distance $r_\del \approx $ 	0.335, 0.5, and 0.775	
for the Lucy, smooth, and Shell weighting functions, respectively.

For the case of $N = $ 2 particles, the SG potential can be expressed in terms of a central pair potential, $U(\bR) = \tUtwo(\Rtwo)$, with 
\begin{equation}
\label{eq-tUtwo-sg}
\tUtwo(r) = 2 \Udel\!\left(\overline{w}(r) \right) \left | \bar{w}'(r) \right|^2 	,
\end{equation}
which is plotted in Fig.~\ref{fig:2PartSGonly}c.
By stabilizing local density gradients, the SG potential generates an attraction between the pair.
The corresponding force function is 
\begin{equation}
\label{eq-tFtwo-sg}
\tFtwo(r) 
	= 
		- \dd \tUtwo(r) / \dd r 
	= 
		 2 \bar{w}'(r) \left [ 
		 			\Fdel\!\left(\wbar\right) \left|\bar{w}'(r)\right|^2 
				-
					2 \Udel\!\left(\wbar\right) \bar{w}''(r) 
				\right ]
	.
\end{equation}
Note that Eq.~\eqref{eq-tFtwo-sg} includes terms that depend upon both  $\Udel$ and $\Fdel = - \dd \Udel / \dd\rhoL$ in Eq.~\eqref{eq-tFtwo-sg}.
Appendix 2 demonstrates that, for $N = 2$, one can shift $\Udel(\rhoL) \to \widetilde U_\del(\rhoL) = \Udel(\rhoL) + c \left| \del \rhoL \right|^{-2}$ (for an arbitrary constant $c$) without impacting the simulated forces. 

Figure~\ref{fig:2PartSGonly}c demonstrates that the SG potential depends very strongly upon the shape of the weighting function. 
In particular, the potential for the DPD weighting function is always attractive and becomes increasingly attractive as particles approach.
The potential for the sphere weighting function is also always attractive but much weaker. 
In contrast, the potentials for the Lucy, smooth, and Shell weighting functions switch from attractive to repulsive near the distance, $\rdel$, that maximizes $\left| \del \rhoL \right|^2.$

The solid curves in Figs.~\ref{fig:2PartSGonly}b and \ref{fig:2PartSGonly}d present, for each weighting function, the analytically determined, canonical probabilities for local densities,  $\Pb(\rhoL)$, and for pair distances, $P_{\rm r}(r)$, respectively.
These distributions are consistent with the preceding discussion.
The distance and LD distributions for the DPD weighting function are peaked near $\rdel$ = 0 and the corresponding local density, $\rhomax$, respectively.
The distance distributions for the Lucy, smooth, and Shell weighting functions are all peaked near the corresponding distances $\rdel > 0$, while the LD distributions are peaked near the corresponding local density, $\rhoLd = \bar{w}(\rdel)$.
Because $\left| \del \rhoL \right|^2$ varies less rapidly with distance for the sphere weighting function, the corresponding distributions are much broader.

The dashed curves in Figs.~\ref{fig:2PartSGonly}b and \ref{fig:2PartSGonly}d present the distributions sampled by Langevin dynamics simulations in LAMMPS with the potential, $U(\bR)$, defined by Eqs.~\eqref{def-U2b} and \eqref{def-Udel} for the 5 different weighting functions.
The quantitative agreement between the canonical and simulated distributions again illustrates the accuracy of the LAMMPS implementation of SG potentials.

We employed BOCS to determine the simulated potentials from the sampled forces via variational force-matching.
The dashed curves in Fig.~\ref{fig:2PartSGonly}a and \ref{fig:2PartSGonly}c present the SG coefficient functions and pair potentials obtained from independent force-matching calculations with the corresponding basis functions. 
We again observe that BOCS determines the simulated potentials with quantitative accuracy. 
The insets of these panels present more quantitative assessment of these errors.
The inset of Fig.~\ref{fig:2PartSGonly}a reports the error in the net forces generated by the calculated SG coefficient function. 
This error is very small, fluctuates about 0, and reflects the relationship between $\Udel$ and $\Fdel$.
The inset of Fig.~\ref{fig:2PartSGonly}b demonstrates that the error in the calculated pair force is similarly small, fluctuates about 0, and reflects limited sampling and the boundary conditions placed upon the spline calculations. 

\FloatBarrier

\subsubsection{Test 3: 125 molecules with LD and SG potentials}
\label{subsubsec-Results-Test3}
\begin{figure}[h!]
\ifShowFig
	\includegraphics[width=3.25in]{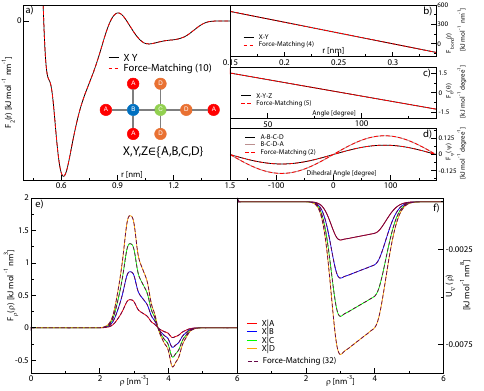}
\fi
	\caption{
	Forces governing the third test.
	Panel a presents the pair force function describing all nonbonded pair interactions, while the inset illustrates the connectivity of each molecule.
	Panels b, c, and d present the bond, angle, and dihedral force functions, respectively.
	Panels e and f present the LD force functions and SG coefficient functions, respectively. 
	The solid black curves present the standard (i.e., ground truth) functions defined by the simulated potential, while the dashed colored curves present the corresponding functions calculated via variational force-matching.
	}
	\label{fig:fullffreco}
\end{figure}

Our third test considers a much more complex system of 125 molecules.
Each molecule consists of 9 particles of 4 distinct types ($\tau = $ A, B, C, or D) that are connected in a branched topology, as illustrated in the inset of Fig.~\ref{fig:fullffreco}.
The B and C sites are located at the two branch points, while the A and D sites are located at the surface of the molecule. 

The bonded potential included 4 distinct bond potentials to govern the A-B, B-C, C-D, and D-A bonds; 5 distinct angle potentials to govern the A-B-A, A-B-C, B-C-D, C-D-A, and D-C-D angles; and 2 distinct dihedral potentials to govern the A-B-C-D and B-C-D-A dihedrals.
Given the four distinct site types, the nonbonded potential was defined by 10 central pair potentials, 16 LD potentials, and 16 SG coefficient functions.
In addition to governing intermolecular interactions, the nonbonded potential also governed intramolecular 1-5 interactions between A sites on opposite sides of the molecule, i.e., we excluded intramolecular 1-2, 1-3, and 1-4 interactions from the nonbonded potential.

As described in Section~\ref{subsubsec-Methods-Test3}, we defined the 53 distinct potentials with 13 different functions in order to simplify the presentation. 
Figures~\ref{fig:fullffreco}b and \ref{fig:fullffreco}c present the linear functions that defined the bond and angle forces, respectively, while Fig.~\ref{fig:fullffreco}d presents the dihedral angle force functions.
Figure~\ref{fig:fullffreco}a presents the function that was used to define all nonbonded pair forces. 
This pair force includes a repulsive hard core region for $r < $ 0.507~nm, 
as well as attractive regions over the intervals $0.507\text{~nm} < r < 0.863$~nm and $0.950 \text{~nm}< r < 1.50$~nm.
Moreover, while this model treated 16 distinct types of local density, we adopted the same Lucy weighting function with $\rc = $ 0.65~nm to define the local density, $\rho_{\tau|I}$, of $\tau$ type sites around each site, $I$. 
Similarly, we assumed that the LD potentials, $U_{\rho \tau|\tau_I}(\rho_{\tau|I})$, and SG coefficient functions, $U_{\del \tau|\tau_I}(\rho_{\tau|I})$, depended only upon the type, $\tau_I$, of the central site, $I$, and not upon the type, $\tau$, of the surrounding sites.
Figures~\ref{fig:fullffreco}e and \ref{fig:fullffreco}f present the LD force functions and SG coefficient functions, respectively. 
The LD potentials have a single minimum at $\rhoL = $ 3.7~\invnm~and only generate forces in the interval, 1.63~\invnm~$ < \rhoL < $ 5.35~\invnm.
The SG coefficient functions stabilize density gradients when 1.91~\invnm~$ < \rhoL < $ 5.44~\invnm~and vanish outside of this interval.
Furthermore, the LD potential and SG coefficient functions for different site types are related by scalar multiples.

\begin{figure}[h!]
\ifShowFig
	\ifShowBigFig
		\includegraphics[width=5in]{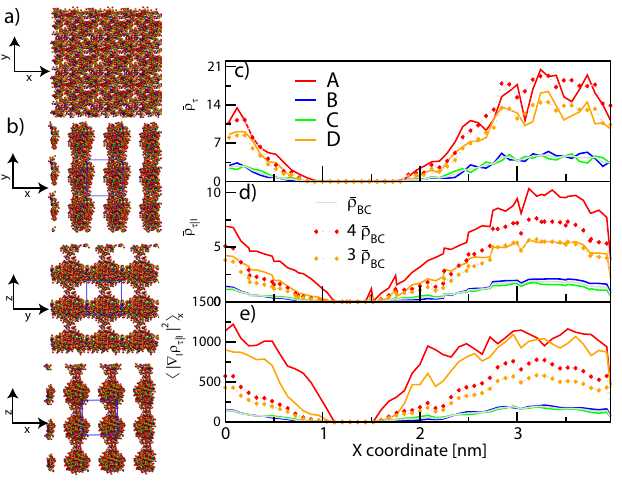}
	\fi
\fi
\caption{
Analysis of simulated morphology.
Panel a presents a visualization of the initial configuration, while panel b presents visualizations of the configuration after 6~ns of simulation. 
Panels c -- e characterize the average density of $\tau = $ A, B, C, and D sites as a function of the x-coordinate in the simulation box. 
Specifically, 
panel c presents the average total density, $\brho_{\tau}(x)$, of $\tau$ sites; 
panel d presents the average local density, $\brho_{\tau|I}(x)$, of $\tau$ sites around each site, $I$, of type A; 
and 
panel e presents the average square gradient of the $\tau$ local density, 
$\llangle |\dI \rho_{\tau|I} |^2 \rrangle_x$, 
around each site, $I$, of type A. 
Panels c, d, and e report densities and square gradients in units of c) \invnm; d) \invnm; and  e)$\text{nm}^{-8}$.
The light gray curves in these panels average the corresponding curves for the B and C sites.
The dotted red and orange curves in these panels scale this gray curve by factors of 4 and 3 in order to account for the multiplicity of A and B sites, respectively.  
	}
	\label{fig:fullffsimulation}
\end{figure}

We employed LAMMPS to simulate the system at a constant temperature, $T = $ 350~K, in a fixed cubic box with sides of length, $L = $ 4~nm, and periodic boundary conditions in all 3 directions.
Starting from the random homogeneous configuration shown in Fig.~\ref{fig:fullffsimulation}a, the system quickly evolved to form the porous two dimensional sheet shown in Fig.~\ref{fig:fullffsimulation}b that is roughly parallel to the y-z plane.
The SI demonstrates that the simulated morphology reflects the periodic boundary conditions, as the 125 molecules form an approximately spherical drop when simulated in a sufficiently large volume.

The remainder of Fig.~\ref{fig:fullffsimulation} characterizes the spatial distribution of sites transverse to the x-coordinate of the simulated box.
The average partial densities, $\brho_{\tau}(x)$, in Fig.~\ref{fig:fullffsimulation}c indicate the vacuum region separating successive periodic sheets and largely follow the multiplicity of the various site types. 
The average local densities, $\brho_{\tau|I}(x)$, in Fig.~\ref{fig:fullffsimulation}d demonstrate that A sites preferentially associate with other A sites and tend to localize at the surface of the sheets, while B and C sites primarily localize in the interior. 
The average square gradients, $\llangle |\dI \rho_{\tau|I} |^2 \rrangle_x$, around sites, $I$, of type A in Fig.~\ref{fig:fullffsimulation}d demonstrate that A and D sites both localize in regions of relatively high density gradients.

\ShowBigFigtrue

We then employed BOCS to determine the simulated force functions from the net forces sampled by the LAMMPS simulation.  
This variational force-matching calculation represented the bond, angle, and dihedral force functions with 19,152 cubic spline polynomial functions, while representing the pair force, LD force, and SG coefficient functions with 5,957 cubic spline functions. 
Importantly, this calculation treated each of the 11 bonded potentials and each of the 42 nonbonded potentials as distinct.
For example, we calculated all 10 distinct types of nonbonded pair force functions and did not assume that they were all defined by the same function.

The dashed curves in Fig.~\ref{fig:fullffreco} present the calculated 53 force functions. 
BOCS determines the 26,348 potential parameters  very accurately.
Moreover, the force-matching calculation is quite robust with respect to statistical sampling.
For instance, the SI demonstrates that the simulation rarely sampled configurations with 4.5~\invnm  $ \le \rho_{B|I}, \rho_{C|I} \le $ 6~\invnm.
Nevertheless, BOCS accurately determined the corresponding LD force and SG coefficient functions in this range.

\FloatBarrier

\newcommand{\rholiq}{\rho_{\rm liq}}
\newcommand{\rhovap}{\rho_{\rm vap}}
\newcommand{\Pext}{P_{\rm ext}}

\newcommand{\rhohi}{\rho_{\rm hi}}
\newcommand{\rholo}{\rho_{\rm lo}}

\subsection{Transferable 1-site water model }
Finally, we briefly illustrate the promise of LD potentials for improving the structural fidelity, thermodynamic properties, and transferability of bottom-up CG models. 
Here we adopt the SPC/E as a reasonably accurate all-atom (AA) water model.\cite{Berendsen:1987}
We consider CG models that represent each water molecule by a single site located at its mass center.

\begin{figure}[h!]
\ifShowFig
	\includegraphics[width=5.0in]{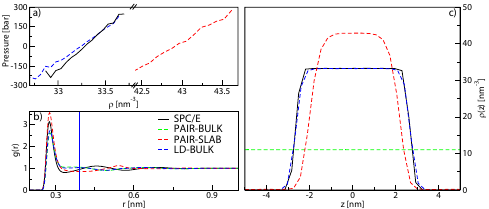}
\fi
	\caption{
	Characterization of SPC/E and CG water models. 
	Panels a and b report the pressure-density equations of state and the simulated rdf's, respectively, obtained from constant NPT bulk simulations at a temperature $T = $ 300~K and an external pressure $\Pext = $ 1~bar. 	
	Panel c reports the simulated density profile transverse to the z-coordinate  from constant NVT slab simulations at a temperature $T = $ 300~K.
	The solid black curves present results for the SPC/E model, while the  dashed green, red, and blue curves present results for the PAIR-BULK, PAIR-SLAB, and LD-BULK models, respectively. 
	Panel a does not include the pressure-density equation of state for the PAIR-BULK model because this model vaporized under the simulated conditions.
	Accordingly, the green curve in panel b reports the rdf obtained from a constant NVT simulation of the PAIR-BULK model performed at the average volume of the SPC/E constant NPT simulation.
	The vertical blue line in panel b indicates the local density length-scale, $\rc = 0.392$~nm. 
	}
	\label{fig:waterfig}
\end{figure}

The solid black curves in Figs.~\ref{fig:waterfig}a and \ref{fig:waterfig}b report the pressure-density equation of state and the mapped radial distribution function (rdf) obtained from a constant NPT AA simulation of bulk water at a temperature $T = $ 300~K and an external pressure $\Pext = $ 1~bar.
The mapped rdf features a sharp first peak at $r \approx $ 0.276~nm that corresponds to hydrogen-bonding interactions, as well as a broader peak near $r \approx $ 0.458~nm that corresponds to the second solvation shell. 
The solid black curve in Fig.~\ref{fig:waterfig}c reports the interfacial density profile obtained from a constant NVT AA simulation of a liquid slab in coexistence with vapor at $T = $ 300~K.
The slab simulation forms a very sharp interface between a liquid phase with $\rholiq = $ 33.27~\invnm\, and a vapor phase with $\rhovap = $ 0.00018~\invnm.

We employed BOCS to parameterize three different 1-site CG models via variational force-matching.
The PAIR-BULK and PAIR-SLAB models described interactions with conventional pair potentials that were parameterized from the bulk and slab SPC/E simulations, respectively.
Conversely, the LD-BULK model described interactions with both pair and LD potentials that were parameterized from the bulk water simulation.
The LD-BULK model defined the local density by a Lucy weighting function with a length-scale $\rc = $ 0.392~nm that occurs early in the second solvation shell.

\begin{figure}[h!]
\ifShowFig
	\includegraphics[width=5.0in]{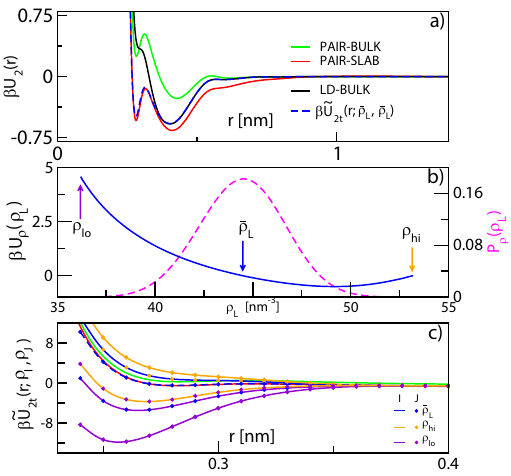}
\fi
	\caption{
	Interaction potentials for CG water models.
	The solid curves in panels a and b present the calculated pair potentials and the calculated LD potential for the BULK-LD model, respectively. 
	The dashed magenta curve in panel b presents the local density distribution sampled by the bulk SPC/E simulation at the temperature $T = $ 300~K and external pressure $\Pext = $ 1~bar. 
	The blue arrow in panel b indicates the mean, $\brhoL$, of this distribution, while the purple and orange arrows indicate the minimum, $\rholo$, and maximum, $\rhohi$, local densities sampled by this SPC/E simulation.	
	Panel c analyzes the influence of the local densities, $\rho_I$ and $\rho_J$, of molecules $I$ and $J$, respectively, upon the corresponding active pair potential, $\tUtt(r;\rho_I,\rho_J)$.
	In panel c, the purple, blue, and orange curves correspond to $\rhoI = \rholo, \brhoL,$ and $\rhohi$, respectively, while the purple, blue, and orange symbols correspond to $\rhoJ = \rholo, \brhoL,$ and $\rhohi$, respectively.
	The green and red curves in panel c reproduce the PAIR-BULK and PAIR-SLAB pair potentials from panel a, respectively.
	The dashed blue curve in panel a presents the active pair potential for $\rho_I = \rho_J = \brhoL$. 
	}
	\label{fig:activeLDforces}
\end{figure}

The solid curves in Fig.~\ref{fig:activeLDforces}a present the calculated pair potentials for the 3 CG models.
Each pair potential features a very narrow minimum at $r \approx$ 0.28~nm that corresponds to hydrogen-bonding interactions, as well as a broader global minimum that occurs at $r \approx$ 0.41~nm near the second peak of the mapped rdf.  
The PAIR-BULK potential is very repulsive and also acts over a very short range.
The PAIR-SLAB potential is not only more attractive, but also acts over a much longer distance range.
Interestingly, the calculated LD-BULK pair potential is very similar to the PAIR-BULK pair potential at both short, $r < $ 0.27~nm, and at long range, $r > $ 0.6~nm.
However, the LD-BULK pair potential is much more attractive over the intermediate regime, 0.27~nm $ < r < $ 0.6~nm.
In particular, the global minima of the PAIR-SLAB and LD-BULK pair potentials are quite similar. 
Nevertheless, none of the minima in the calculated pair potentials are deeper than 1~$\kT$.

The solid blue curve in Fig.~\ref{fig:activeLDforces}b presents the calculated LD potential for the LD-BULK model.
The dashed magenta curve in Fig.~\ref{fig:activeLDforces}b presents the distribution of local densities, $\rhoL$, sampled by the bulk SPC/E simulation that was employed to parameterize the LD-BULK model.
Because we include the self-term in defining the local density, the sampled local densities are significantly greater than the equilibrium bulk density.
Moreover, because it is defined over such a short length-scale, the local density distribution is much broader than the simulated bulk density distribution.
The calculated LD potential is attractive at most sampled local densities, including the average local density, $\brhoL$.
However, the LD potential becomes slightly repulsive at the highest local densities.

While the BULK-LD model generates pair-additive forces, these pair forces are modulated by the local environment of each pair.
Specifically, the total force, $F_{2t}$, between each pair, $(I,J)$, is directed along the vector between the pair and may be expressed
\begin{equation}
\label{F-2-total}
F_{2t}(\RIJ, \rhoI,\rhoJ) 
	= 
		F_2(\RIJ) 
	+ 
		\left[ \Frho(\rhoI) + \Frho(\rho_J) \right] \bar{w}'(\RIJ)	,
\end{equation}
where $F_2(r) = - \dd U_2(r) / \dd r$ is the pair force function, while $\Frho(\rho) = - \dd \Urho(\rho) / \dd \rho$ is the LD force function.
Thus, the total pair force depends upon the pair distance, $\RIJ$, as well as the local density, $\rhoI$ and $\rho_J$, of molecules $I$ and $J$.
Accordingly, we define an ``active,'' context-dependent pair potential, $\tUtt(r;\rhoI,\rhoJ)$, for the BULK-LD model by integrating $-F_{2t}(r,\rhoI,\rhoJ)$ with respect to $r$, while holding $\rhoI$ and $\rhoJ$ constant.
Because $\bar{w}(r) = 0 $ for $r > \rc$, the local density only impacts $\tUtt(r;\rhoI,\rhoJ)$ between pairs that are separated by $r < \rc$.
We note that the total potential of the BULK-LD model cannot be expressed in terms $\tUtt(r;\rhoI,\rhoJ)$.
Nevertheless, $\tUtt(r;\rhoI,\rhoJ)$ provides useful qualitative insight into the impact of the environment upon pair interactions in the PAIR-LD model.

Figure~\ref{fig:activeLDforces} illustrates the influence of the local environment upon the active pair potential, $\tUtt(r;\rhoI,\rhoJ)$, for $r < \rc$.
The dashed blue curve in Fig.~\ref{fig:activeLDforces}a presents $\tUtt(r;\brhoL,\brhoL)$, which describes the interaction when both particles experience the average local density of the SPC/E model, i.e., $\rhoI = \rhoJ = \brhoL$.
In this case, the active pair potential is very similar to the PAIR-SLAB potential in the first two solvation shells, but is very similar to the PAIR-BULK potential beyond the second solvation shell. 
Moreover, Fig.~\ref{fig:activeLDforces}c demonstrates that the active pair potential becomes even much more attractive as the local density decreases.
This effect is especially pronounced in the first solvation shell as the hydrogen-bonding minimum not only deepens but also shifts to shorter distances with decreasing $\rhoL$.
In particular, if particle $I$ experiences the average local density, $\rho_I = \brhoL$, while particle $J$ experiences the lowest local density, $\rho_J = \rholo$, then the hydrogen-bonding minimum of the corresponding active pair potential is approximately $-6 \kT$.
In this regime, the active potential is much more attractive than the PAIR-SLAB potential.
Furthermore, if both particles are in a low density environment, $\rhoI = \rhoJ = \rholo$, then the hydrogen-bonding minimum deepens to approximately $-12 \kT$.
Conversely, when both particles are in a high density environment, $\rhoI = \rhoJ = \rhohi$, then the hydrogen-bonding ``minimum'' is approximately +1$\kT$ and not even locally stable. 
In this case, the active potential is significantly more repulsive than the PAIR-BULK potential.

The dashed red and blue curves in Figs.~\ref{fig:waterfig}a and b report the results of constant NPT simulations with the PAIR-SLAB and LD-BULK models, respectively.
Because the PAIR-BULK potential is almost entirely repulsive, this model dramatically overestimates the internal pressure and simply vaporizes at 1~bar external pressure.
Conversely, because the PAIR-SLAB potential is much more attractive, it stabilizes a condensed phase at 1~bar but overestimates the density and compressibility of the SPC/E model by 28.9~\% and 37.6~\%, respectively.
As a consequence of this significantly enhanced density, the PAIR-SLAB model also describes the pair structure quite poorly and, in particular, significantly overestimates the first peak of the SPC/E rdf.
The LD-BULK model reproduces the SPC/E equation of state much more accurately, underestimating the SPC/E density by only 0.1~\% and overestimating the SPC/E compressibility by 27.4~\%.
Unfortunately, the LD-BULK model generally underestimates the order within the first solvation shells of the SPC/E model.
For comparison, the solid green curve in Fig.~\ref{fig:waterfig}b reports the rdf obtained from constant NVT simulations of the PAIR-BULK model at the correct density.
The inclusion of the LD potential slightly improves the agreement with the SPC/E rdf in the first solvation shell. 
Nevertheless, the 
similarity between the dashed blue and green curves indicates that the 
LD potential exerts relatively little influence upon the local water structure.

Finally, the dashed, colored curves in Fig.~\ref{fig:waterfig}c report the density profiles obtained from constant NVT slab simulations with the three CG models. 
Because the PAIR-BULK model cannot stabilize the liquid phase at 1~bar pressure, this model forms a uniform homogeneous phase. 
Conversely, the PAIR-SLAB model stabilizes liquid-vapor coexistence but significantly overestimates the density of the liquid phase, as already indicated in Fig.~\ref{fig:waterfig}b.
Moreover, although it was parameterized from the SPC/E slab simulation, the PAIR-SLAB model badly overestimates the width of the liquid-vapor interface.
In contrast, although it was parameterized from a bulk simulation, the LD-BULK model  accurately reproduces the densities of the coexisting phases model. 
Moreover, the LD-BULK model also quite reasonably reproduces the width of the liquid-vapor interface.

We refer readers to Ref.~\citenum{Lesniewski:2026a} for a much more detailed analysis of LD potentials for 1-site water models. 
In particular, we employed variational force-matching to parameterize pair and LD potentials based upon a single bulk simulation of the TIP4P water model.\cite{jorgensen_comparison_1983}
Despite providing a rather poor description of the TIP4P rdf, this CG model reproduced cooperative hydrophobic phenomena quite accurately. 
Thus, LD potentials significantly improve the thermodynamic properties of CG models for bulk systems, their structural fidelity in describing inhomogeneous systems, and their transferability between bulk liquid and vapor phases, as well as between bulk and interfacial environments.

\FloatBarrier
\section{Conclusions}
\label{sec-concl}
Since its original release,\cite{dunn_bocs:_2018} BOCS has provided an accurate and robust software package for parameterizing bottom-up CG models of soft materials.
In addition to implementing the force-matching variational principle for parameterizing molecular mechanics potentials,\cite{Izvekov:2005d,Izvekov:2005e,Noid:2008a,Noid:2008b} BOCS provided several unique capabilities.
For instance, BOCS implemented a generalized-Yvon-Born-Green theory for relating these potentials to structural correlation functions,\cite{Noid:2007a,Rudzinski:2015bb,Mullinax:2009b,Mullinax:2010,Rudzinski:2012vn,Rudzinski:2015bb} a simple framework for improving their transferability across multiple systems,\cite{Mullinax:2009a} and a self-consistent pressure-matching method for accurately reproducing pressure-density equations of state.\cite{Dunn:2015wn}

In this work, we have described the latest release of BOCS for parameterizing LD and SG potentials via variational force-matching. 
This release is compatible with complex molecular topologies and implements a wide range of weighting functions for defining the local density.
In addition, we have described a significant extension of the PKG-BOCS package for simulating these LD and SG potentials in LAMMPS. 
This package implements flexible spline-based representations of these potentials, as well as a library of functional forms that require fewer parameters.
We have been collaborating with LAMMPS developers to incorporate the PKG-BOCS package and anticipate it will be included with the next LAMMPS release.
However, in the interim, we are publicly distributing a version of the PKG-BOCS package configured with LAMMPS April 2026 on our Github site\cite{GitHubBocs} along with all the requisite documentation for its use.
Moreover, we are distributing Doxygen documentation and tutorials for using this software in BOCS and LAMMPS.

In addition to describing this new software, we have also reported calculations illustrating its use.
These calculations have provided physical insight into the forces generated by LD and SG potentials.
In particular, we have highlighted the impact of the weighting function used to define the local density.
Moreover, we have illustrated the practical benefits of these LD potentials for improving the structural fidelity, thermodynamic properties, and transferability of bottom-up CG models. 
Most importantly, our calculations demonstrate that our implementation of LD and SG potentials is both accurate and robust.

The present work also indicates several directions for future studies. 
For instance, the accuracy of LD potentials determined via force-matching depends quite strongly upon the length-scale, $\rc$, defining the local density. 
It would be very helpful to gain physical insight for predicting the optimal $\rc$.
Moreover, future studies should investigate the use of LD potentials in CG models with long-ranged electrostatics. 
Nevertheless, we hope that this new version of BOCS and the PKG-BOCS package in LAMMPS may prove useful for developing bottom-up CG models of complex soft materials.

\FloatBarrier

\section{Appendices}

\newcommand{\tI}{{\tau_I}}
\newcommand{\tJ}{{\tau_J}}
\newcommand{\cStI}{\cS_{\tau|I}}
\newcommand{\rhotI}{\rho_{\tau|I}}
\newcommand{\ctau}{c_\tau}
\newcommand{\cttI}{c_{\tau|\tI}}
\newcommand{\RIK}{R_{IK}}
\newcommand{\hbRIK}{\hbR_{IK}}

\newcommand{\psizl}{\psi_{\zt\la}}
\newcommand{\Urz}{U_{\rm R0}}
\newcommand{\bFIz}{\bF_{I;\rm R0}}

\subsection{Appendix 1: Basis functions for variational force-matching.}
The force field basis vectors associated with bonded and pair non-bonded potentials are easily derived. 
We define $\Urz$ as the corresponding part of the total interaction potential:
\begin{equation}
\Urz(\bR) \equiv \sum_\zt \sum_\la \Uzt(\psizl(\bR))	.
\end{equation}
Here $\zt$ indicates a bonded or pair non-bonded interaction, $\Uzt$ is the corresponding potential that is a function of a corresponding scalar variable, $\psizl(\bR)$, where $\la$ indicates the particular instance of the interaction. 
For each of these potentials, we define a corresponding force function, $\Fzt(x) = - \dd \Uzt(x) / \dd x$, that we represent with a set of basis functions, $\fzd(x)$:
\begin{equation}
\Fzt(x) = \sum_d \phizd \fzd(x) ,
\end{equation}
where $\phizd$ are the corresponding force field coefficients.
The total force from $\Urz$ may then be expressed:
\begin{equation}
\bFIz(\bR) \equiv - \dI \Urz(\bR) = \sum_{\zt}\sum_d \phizd \cGIzd(\bR)	,
\end{equation}
where the force field basis vector associated with $\fzd$ is 
\begin{equation}
\cGIzd(\bR) \equiv \sum_\la \fzd(\psizl(\bR)) \dI \psizl(\bR) 		.
\end{equation}

The basis functions associated with LD and SG potentials are more complex. 
Here we derive these basis vectors for the general case that the CG model treats multiple distinct types of CG sites.
We first assign a type, $\tI \in \{ \al, \bt, \ga, \cdots \}$, for each particle, $I = 1, \ldots, N$.
This allows us to partition the set, $\cSI$, of particles that interact with site $I$ via non-bonded interactions into contributions from each different type of site:
\begin{equation}
\cSI = \bigcup_\tau \cStI	,
\end{equation}
where $\cStI = \{ J | \tau_J = \tau \text{ and $J$ interacts with $I$ via nonbonded potentials} \}$. 
Note that $I \notin \cSI$. 
Moreover, the following are all equivalent: (1) $J \in \cSI$;  (2) $J \in \cS_{\tau_J|I}$; (3) $I \in \cS_J$; and (4) $I \in \cS_{\tI|J}$.
For each ordered pair of particle types, $(\tau_2|\tau_1)$, we define a corresponding weighting function, $\wb_{\tau_2|\tau_1}(r)$.
This allows us to define the local density, $\rhotI$, of $\tau$ sites around particle $I$:
\begin{equation}
\rhotI 
	\equiv 
		\ctau \de_{\tau\tau_I} 
		+ 
		\sum_{J\in\cStI} \wb_{\tau|\tI}(\RIJ) ,
\end{equation}
where $\ctau$ is a configuration-independent constant that depends upon $\tau$ and  accounts for any self-contribution to the local density since $I \notin \cStI$.
Note that 
\begin{equation}
\dI \rho_{\tau|J} 
	= 
				\de_{I,J} \sum_{K\in\cStI} \wb'_{\tau|\tI}(\RIK) \hbRIK
	+ 	
				\te[I\in\cS_{\tau|J}] \wb'_{\tI|\tJ}(\RIJ) \hbRIJ	,
\end{equation}
where $\te[C]$ is an indicator function that equals 1 when the condition $C$ is satisfied and otherwise vanishes.
In particular, $\te[I\in\cS_{\tau|J}] = \de_{\tI,\tau} \te[I\in \cS_J]$.
(Note that $\te[I\in\cS_J] = 0$ if $I = J$.)

\newcommand{\Uttp}{U_{\tau_2|\tau_1}}
\newcommand{\Fttp}{F_{\tau_2|\tau_1}}

We  define the LD potential:
\begin{equation}
\Uld(\bR) = \sum_{J=1}^N \sum_\tau U_{\tau|\tJ}(\rho_{\tau|J})	.
\end{equation}
We define a force function, $\Fttp(\rho) = - \dd \Uttp(\rho)/\dd\rho$, for each LD potential, $\Uttp$.
We represent $\Fttp(\rho)$ with a linear combination of basis functions:
\begin{equation}
\Fttp(\rho) = \sum_d \phi_{\tau_2|\tau_1;d} \: f_{\tau_2|\tau_1;d}(\rho)	.
\end{equation}
This then defines the basis set for the LD potential:
\begin{equation}
\bF_{I;\rm LD}(\bR) 
	\equiv
		- \dI \Uld(\bR) 
	= 
		\sum_{\tau_2} 
			\sum_{\tau_1} 
				\sum_d		
						\phi_{\tau_2|\tau_1;d} \; \cG_{I;\tau_2|\tau_1 d}(\bR)		
\end{equation}
where the force field basis vector, $\cG_{I;\tau_2|\tau_1 d}(\bR)$, associated with $\Uttp$ may be expressed
\begin{eqnarray}
\nonumber
\cG_{I;\tau_2|\tau_1 d}(\bR) 
	& = & 
			\de_{\al\tau_1}	
						f_{\tau_2|\al; d}(\rho_{\tau_2|I}) 
				\sum_{J\in\cS_{\tau_2|I}} 
						\wb'_{\tau_2|\al}(\RIJ) \hbRIJ			
						\\
\label{def-cGIttp}
	& & 
		+
			\de_{\al\tau_2}	
				\sum_{J\in\cS_{\tau_1|I}} 
						f_{\al|\tau_1; d}(\rho_{\al|J}) 
						\wb'_{\al|\tau_1}(\RIJ) \hbRIJ			,
\end{eqnarray}
where we have set $\al \equiv \tau_I$ to slightly simplify the expression.

\newcommand{\deIJK}{\de_{I|JK}}
\newcommand{\LtKJij}{L_{\tau K|J ij}}
\newcommand{\RJK}{R_{JK}}
\newcommand{\bAtJ}{\bA_{\tau|J}}
\newcommand{\AtJj}{A_{\tau|Jj}}
\newcommand{\rhotJ}{\rho_{\tau|J}}
\newcommand{\bB}{{\mathbf{B}}}
In order to define the basis vectors for the SG potential, we first define $\bAtJ = \nabla_J \rhotJ = (\AtJj) $ with $\AtJj = \nabla_{Jj} \rhotJ = \ptl \rhotJ / \ptl R_{Jj}$, where the lower case letters indicate Cartesian directions, $i, j = 1, 2, 3$.
We also need to evaluate 
\begin{equation}
\frac{\ptl}{\ptl R_{Ii}} \AtJj
	= 
		\sum_{K\in\cS_{\tau|J}}	
			\deIJK \LtKJij ,
\end{equation}
where $\deIJK = \de_{IJ} - \de_{IK}$ and we have also introduced the matrix, 
$L_{\tau K|J} = \left( \LtKJij \right)$ ,
with elements
\begin{equation}
\LtKJij
	=
		f_{\tau|J}(\RJK) \de_{ij}
	+
		g_{\tau|J}(\RJK) 
			\hbR_{JKi}\hbR_{JKj}	,
\end{equation}
where
\begin{eqnarray}
f_{\tau|J}(r) 	& = & r\inv \wb'_{\tau|\tau_J}(r) 	\\
g_{\tau|J}(r) 	& = & \wb''_{\tau|\tau_J}(r) - f_{\tau|J}(r) 
\end{eqnarray}
and $\hbR_{JKi}$ is the $i$\tsup{th} Cartesian component of the unit vector, $\hbRJK$.
This matrix allows us to express 
\begin{eqnarray}
\frac{\ptl}{\ptl \bRI} 
 	\left| \nabla_J \rho_{\tau|J} \right|^2
	& = &  
		 \frac{\ptl}{\ptl \bRI} \bAtJ^2
		=
		2 \sum_{K\in\cS_{\tau|J}}	\deIJK
			\bB_{\tau K | J} 
\end{eqnarray}
where 
\begin{eqnarray}
\bB_{\tau K | J}  
	\equiv 
		L_{\tau K | J}	\bAtJ
	= 
			f_{\tau|J}(\RJK) \bAtJ		
		+ 
			g_{\tau|J}(\RJK)
				\left(\hbRJK \cdot \bAtJ \right) 
				\hbRJK
\end{eqnarray}
\newcommand{\UdtJ}{U_{\nabla \tau|\tau_J}}
\newcommand{\Udtt}{U_{\nabla \tau_2|\tau_1}}
\newcommand{\Fdtt}{F_{\nabla \tau_2|\tau_1}}
\newcommand{\fdttJd}{f_{\nabla \tau_2|\tau_1;d}}
\newcommand{\phidttJd}{\phi_{\nabla \tau_2|\tau_1;d}}
\newcommand{\FdtJ}{F_{\nabla \tau|\tau_J}}
We now define the more general SG potential: 
\begin{equation}
\Usg(\bR) = \sum_{J=1}^N \sum_\tau \UdtJ(\rhotJ) \bAtJ^2	.
\end{equation}
We represent the coefficient function, $\Udtt$, with a simple basis set, which then determines a corresponding representation of $\Fdtt(\rho) = -\dd \Udtt(\rho)/ \dd \rho$:
\begin{eqnarray}
\Udtt(\rho) 
	& = & 
		\sum_{d} \phidttJd \fdttJd(\rho) 	\\
\Fdtt(\rho) 
	& = & 
		- \sum_{d} \phidttJd \fdttJd'(\rho) 			
\end{eqnarray}
This then determines the basis set for the SG potential:
\begin{equation}
\bF_{I;\rm SG}(\bR) 
	\equiv
		- \dI \Usg(\bR) 
	= 
		\sum_{\tau_2} 
			\sum_{\tau_1} 
				\sum_d		
					\phi_{\del \tau_2|\tau_1;d} \; 
						\cG_{I;\del\tau_2|\tau_1 ;d}(\bR)		
\end{equation}
where
\newcommand{\bG}{{\mathbf{G}}}
\newcommand{\bH}{{\mathbf{H}}}
\begin{eqnarray}
\cG_{I;\del\tau_2|\tau_1 ;d}(\bR)
	& = & 
		\cG^{(1)}_{I; \del \tau_2|\tau_1; d}
	+
		\cG^{(2)}_{I; \del \tau_2|\tau_1; d}	
\end{eqnarray}
and
\begin{eqnarray}
\cG^{(1)}_{I; \del \tau_2|\tau_1; d}
	& = & 
	-\de_{\al\tau_1}
		\!\!
		\sum_{J\in \cS_{\tau_2|I}}
		\!\!\!\!
		\left\{
			f'_{\del\tau_2|\al; d}(\rho_{\tau_2|I})
				\bA_{\tau_2|I}^2 \wb'_{\tau_2|\al}(\RIJ) \hbRIJ
		+ 
			2 f_{\del\tau_2|\al; d}(\rho_{\tau_2|I}) \bB_{\tau_2 J|I}
		\right\}
		\\
\cG^{(2)}_{I; \del \tau_2|\tau_1; d}	
	& = & 
		- \de_{\al \tau_2}
		\!\!
		\sum_{J\in \cS_{\tau_1|I}}
		\!\!
		\left\{
			f'_{\del\al|\tau_1; d}(\rho_{\al|J})
				\bA_{\al|J}^2 \wb'_{\al|\tau_1}(\RIJ) \hbRIJ
		- 
			2 f_{\del\al|\tau_1; d}(\rho_{\al|J}) \bB_{\al I|J}
		\right\}
\end{eqnarray}

\subsection{Appendix 2: 2 Particle SG Invariance}
\label{sec:SGINV}
\newcommand{\rhor}{\rho_r}
\newcommand{\rrho}{r_\rho}
The definition of the SG potential, 
$
\Usg(\bR) 
	 \equiv  	
					\sumIN \Udel(\rhoI) \left| \dI \rhoI \right|^2			,
$
in Eq.~\eqref{def-Usg} implies that the forces and the canonical configuration distribution are unchanged when the SG coefficient function, $\Udel(\rhoI)$, is replaced with the generalized coefficient  $\widetilde U_\del(\rhoI,\dI\rhoI) = \Udel(\rhoI) + c \left| \dI \rhoI \right|^{-2}$.
Section~\ref{para-SGpotl} considered a system of $N = 2$ particles interacting via a SG potential, for which $\rhoI = \overline w(R_{12})$ and $\left| \dI \rhoI \right|^2 = |\overline w'(R_{12})|^2$.
In this case, $\rhoI$ and $R_{12}$ are in 1-to-1 correspondence for $0 < R_{12} < r_c$, which implies that $\left| \dI \rhoI \right|^{-2}$ is a function of $\rhoI$.
Thus, for $N = 2$, the generalized coefficient may be expressed as a function of only the local density, i.e., $\widetilde U_\del = \widetilde U_\del(\rhoI)$.
Below we explicitly derive this invariance for $N = 2$.

In the case of $N = 2$ the total potential can be expressed in terms of the effective pair potential, $\tUtwo$, given by Eq.~\eqref{eq-tUtwo-sg} that corresponds to the effective pair force, $\tFtwo$, given by Eq.~\eqref{eq-tFtwo-sg}.
We now consider replacing $\Udel(\rhoI)$ with $\widetilde U_\del(\rhoI) \equiv \Udel(\rhoI) + y(\rhoI)$, where $y(\rhoI)$ is an unknown function.
The equilibrium distribution will remain unchanged as long as the effective pair force remains unchanged. 
By substituting $\widetilde U_\del(\rhoI)$ into Eq.~\eqref{eq-tFtwo-sg}, we see that this invariance is achieved when $y$ satisfies the following differential equation: 
\begin{equation}
	\bar{w}'(r) 
		\left[ 
			y'(\rhor) \left|\bar{w}'(r)\right|^2 + 2 y(\rhor) \bar{w}''(r)  
		\right] 
	=
		0,
		\label{eqn:SG2Invreq-a}
\end{equation}
where $\rhor = \wbar$.
We consider the distance range over which $\bar{w}'(r) < 0$ and assume that $y(\rhor) > 0 $ over this range.
Eq.~\eqref{eqn:SG2Invreq-a} is then equivalent to
\begin{equation}
\label{eqn:SG2Invreq-b}
0
	= 
		\frac{y'(\rhor)}{y(\rhor)} \bar{w}'(r) + 2 \frac{\bar{w}''(r)}{\bar{w}'(r)}
	=
		\frac{\dd}{\dd r} 
			\left\{ 
					\ln \left[ 
								y(\rhor) 
								\left| \bar{w}'(r) \right|^2 
						\right] 
			\right\}  ,
\end{equation}
which is  satisfied when 
$
	y(\rhor)  
			=  
				c 
				\left| \bar{w}'(r) \right|^{-2}
$ 
or, equivalently,
\begin{equation}
\label{eq-soln-y}
y(\rho) 
		= 
				c \left| \bar{w}'(\rrho) \right|^{-2}
,
\end{equation}
where $\rrho$ is the unique distance for which $\bar{w}(\rrho) = \rho$.
Consequently, there exists a one-parameter family of SG coefficient functions, $\widetilde U_\del(\rho;c) = \Udel(\rho) + c \left| \bar{w}'(\rrho) \right|^{-2}$  that generate the same forces on a pair of isolated particles.

\newcommand{\Udelsim}{U_\del^{\rm sim}}
\newcommand{\UdelFM}{U_\del^{\rm FM}}
\newcommand{\tUdelFM}{\widetilde U^{\rm FM}_\del}

If we had represented $\Udel(\rhoL)$ with a flexible spline basis set, then this invariance would have complicated our assessment of the SG test in Section~\ref{para-SGpotl}.
Specifically, this invariance would imply that a family of equivalent SG coefficient functions, $\tUdelFM(\rho;c) = \UdelFM(\rho) +  c \left| \bar{w}'(\rrho) \right|^{-2}$ all minimized the force-matching variational functional, $\chi^2[U]$. 
However, we avoided these complications by representing  $\Udel$ with a less flexible basis set that was not compatible with $y(\rho) = c \left| \bar{w}'(\rrho) \right|^{-2}$.

\section*{Supporting Information}
Additional details of the local density weighting functions and force-matching calculations, as well as additional analysis of the molecular test system.

\section*{Author Contributions}
\noindent
{\bf Maria C. Lesniewski:} Conceptualization (equal); Data curation (lead); Formal analysis (equal); Investigation (lead); Methodology (equal); Software (lead); Validation (lead); Writing - original draft (lead); Writing - review \& editing (supporting)
\\
{\bf Michael R. DeLyser:} Conceptualization (supporting); Formal analysis (supporting); Methodology (equal); Software (supporting);  Writing - review \& editing (supporting) 
\\
{\bf W. G. Noid:} Conceptualization (equal); Formal analysis (equal); Funding acquisition (lead); Investigation (supporting); Methodology (equal); Project administration (lead); Resources (lead); Supervision (lead);  Writing - original draft (supporting); Writing - review \& editing (lead)

\section*{Data Availability}
The simulation results, analysis, and the exact LAMMPS and BOCS source codes described in this work are available from the Penn State DataCommons\cite{DataCommons} at 
\href{https://doi.org/10.26208/PW48-SA96}{https://doi.org/10.26208/PW48-SA96}.\cite{DataCommonsBocs2026}
The BOCS code version 5.0 code that was used in the reported force-matching calculations may be accessed at \href{https://github.com/noid-group/BOCS}{https://github.com/noid-group/BOCS}, along with corresponding documentation.\cite{GitHubBocs}
The version of PKG-BOCS that was used in the reported simulations, along with rst/html style documentation for using it and interpreting the syntax, may be accessed at \href{https://github.com/noid-group/lammps-BOCS2026}{https://github.com/noid-group/lammps-BOCS2026}.
The final version of PKG-BOCS that has been implemented into LAMMPS may be accessed at \href{https://github.com/lammps/lammps/pull/4737}{https://github.com/lammps/lammps/pull/4737}.
The LAMMPS package is distributed under the GPL2 license,  Copyright (2003) Sandia Corporation.  Under the terms of Contract DE-AC04-94AL85000 with Sandia Corporation, the U.S. Government retains certain rights in this software. Available at   \href{https://www.lammps.org/}{https://www.lammps.org} and \href{https://github.com/lammps/lammps}{https://github.com/lammps/lammps}.
.

\section*{Author Declarations}
\subsection*{Conflict of interest}
The authors have no conflicts to disclose.

\section*{Acknowledgments}
The authors express their sincere gratitude to Axel Kohlmeyer for his assistance in implementing the PKG-BOCS package into the main distribution of LAMMPS.
MCL and WGN also gratefully acknowledge financial support from the National Science Foundation (Grant Nos.~CHE-2154433 and ~CHE-2502094).
Portions of this research were conducted with Advanced CyberInfrastructure computational resources provided by The Institute for Computational and Data Sciences at The Pennsylvania State University (http://icds.psu.edu).
Additionally, parts of this research used the Expanse resource at the San Diego Supercomputer Center through allocation TG-CHE170062 from the Extreme Science and Engineering Discovery Environment (XSEDE)\cite{xsede}, which was supported by National Science Foundation grant number TG-CHE170062.
This work used Anvil at Purdue University\cite{song_anvil_2022} through allocations CHE170062 and CHE250182 from the Advanced Cyberinfrastructure Coordination Ecosystem: Services \& Support (ACCESS) program, which is supported by U.S. National Science Foundation grants \#2138259, \#2138286, \#2138307, \#2137603, and \#2138296.\cite{boerner_access_2023}

\bibliography{WorkingBibFile}	

\begin{tocentry} 
\includegraphics{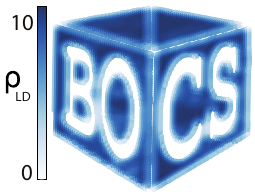}
\end{tocentry}

\end{document}


\begin{abstract}
	This document provides supporting information for the manuscript ``Progress toward a better BOCS: Systematic coarse-graining with local density potentials."
	Section 1 defines the local density weighting functions that have been implemented into BOCS and PKG-BOCS.
	Section 2 details the hyperparameters employed in the reported force-matching calculations.
	Section 3 provides additional analysis of the third test reported in the manuscript.
\end{abstract}

\maketitle

\setcounter{section}{0}

\section{Weighting functions}
\label{sec:weight}
Equation~(6) of the manuscript defines the local density, $\rhoI$, in terms of a normalized weighting function, 
\begin{equation}
	\wbar(r) = \frac{w(r)}{[w]}	,
	\label{eq:wdef}
\end{equation} 
where $w(r)$, is a non-increasing function that is 1 when $r = 0$ and vanishes when $r > \rc$, while $[w]$ is the normalization
\begin{equation}
	[w] = \int_{0}^{r_c} 4 \pi r^2 w(r) dr	.
	\label{eq:wnorm}
\end{equation}
Below we explicitly define the local density weighting functions that we have implemented into BOCS version 5.0 and the PKG-BOCS package in LAMMPS. 
Table~\ref{tab:Indicators} summarizes this information.
While each weighting function vanishes beyond the local density length-scale, $\rc$, the Shell, smooth, and sphere weighting functions also incorporate a second hyperparameter, $r_0$. 
These latter three weighting functions have a constant value of 1 over the interval $0 \le r \le r_0$ and decrease to 0 over the interval $r_0 \le r \le \rc$.
In Fig.~2 of the main text we set $r_0 = \rc / 2$, while in the reported test calculations we set $r_0 = 0$.

\begin{table}[h!]
\centering
\begin{tabular}{|c|c|c|c|}
	
	\hline
	Indicator& hyperparameters & $w(r)$ & $[w]$ \\
	\hline   DPD    & $r_c$        & Eq. \eqref{eq:DPD}       & Eq. \eqref{eq:DPDnorm} \\
	\hline   Lucy   & $r_c$        & Eq. \eqref{eq:Lucy} & Eq. \eqref{eq:Lucynorm} \\
	\hline   Shell  & $r_0$, $r_c$ & Eq. \eqref{eq:shell}  & Eq. \eqref{eq:norm_shell} \\
	\hline   smooth & $r_0$, $r_c$ & Eq. \eqref{eq:smooth} & Eq. \eqref{eq:norm_smooth}  \\
	\hline   sphere & $r_0$, $r_c$ & Eq. \eqref{eq:sphere} & Eq. \eqref{eq:norm_sphere}  \\
	\hline
\end{tabular}
\caption{
Summary of weighting functions that we have implemented in BOCS and the PKG-BOCS package in LAMMPS. 
}
\label{tab:Indicators}
\end{table}

\label{sec:longindicatoreqs}
\subsection{DPD weighting function}
\label{SubSec-w-DPD}
Dissipative particle dynamics (DPD) models\cite{groot_dissipative_1997,Pagonabarraga:2001aa,Warren:2003} often employ the weighting function
\begin{equation}
	w(r) = (1-r/\rc)^2 \theta(1 - r/\rc)	,
	\label{eq:DPD}
\end{equation}
where the Heaviside function $\theta(x) = 0$ for $x < 0 $ and  = 1 for $x > 0$.
The corresponding normalization is
\begin{equation}
	[w] = 2\pi r_c^3/15		.
	\label{eq:DPDnorm}
\end{equation}
The first derivative of the DPD weighting function is continuous everywhere, while the second derivative is discontinuous at $r = \rc$.

\subsection{Lucy weighting function}
\label{SubSec-w-Lucy}
Local density (LD) potentials that are parameterized via force-matching\cite{Moore:2016wn,Wagner:2017xx,Delyser:2019ld,Delyser:2020int,Delyser:2022wn} often define the local density with the Lucy function,\cite{Lucy:1977,monaghan_smoothed_2005}
\begin{equation}
	w(r) = (1-r/r_c)^3(1+3r/r_c) \theta(1 - r/\rc)	.
	\label{eq:Lucy}
\end{equation}
The corresponding normalization is 
\begin{equation}
	[w] = 16\pi r_c^3/105		.
	\label{eq:Lucynorm}
\end{equation}
The first and second derivatives of the Lucy function are continuous for all $ r > 0 $.

\subsection{Shell weighting function}
\label{SubSec-w-Shell}
Sanyal and Shell defined the local density with the following weighting function\cite{Sanyal:2016md,sanyal_transferable_2018}
\begin{equation}
\label{eq:shell}
w(r) = 
\begin{cases}
	1 & r \leq r_0 \\
	d^{-1}(c_0 + c_2 r^2 + c_4 r^4 + c_6 r^6) & r_0 \leq r \leq r_c \\
	0 & r \geq r_c
\end{cases}
\end{equation}
where 
\begin{eqnarray}
	d 	& = & 	(1 - r_0^2/r_c^2)^3, \\
	c_0 	& = & 	(1-3r_0^2/r_c^2), \\
	c_2 	& = &	6r_0^2/r_c^4, \\
	c_4 	& = &	-3(1+r_0^2/r_c^2)/r_c^4,\\
	c_6 	& = & 	2/r_c^6
\end{eqnarray}
The corresponding normalization is 
\begin{equation}
[w] 
	= 
		4 \pi 
		\left\{ 
	   	   \frac{1}{3} r_0^3 
			+
			\frac{1}{d}
			\left[ 
				\frac{1}{3} c_0 (r_c^3 - r_0^3) 
				+ \frac{1}{5}c_2 (r_c^5 - r_0^5) 
				+ \frac{1}{7}c_4 (r_c^7 - r_0^7) 
				+ \frac{1}{9}c_6 (r_c^9 - r_0^9)
			\right]
		\right\}		.
\label{eq:norm_shell}
\end{equation}
The first derivative of the Shell weighting function is continuous everywhere, while the second derivative is discontinuous at $r_0$ and $r_c$.

\subsection{Smooth weighting function}
\label{SubSec-w-smooth}
We have introduced a new weighting function that is also constant over the interior interval $ 0 < r < r_0$ with higher order continuity:
\begin{equation}
\label{eq:smooth}
w(r) = 
\begin{cases}
	1 & r \leq r_0 \\
	d^{-1} \sum_{i=0}^5 c_i r^i & r_0 \leq r \leq r_c \\
	0 & r \geq r_c
\end{cases}
\end{equation}
where the coefficients are
\begin{eqnarray}
d 	& =	& 	(r_0^5 - r_c^5)/120 - (r_0^3 - r_c^3)r_0r_c/24 + (r_0 - r_c)r_0^2r_c^2/12, \\
c_0 	& = 	&	-r_c^5/120 + r_0r_c^4/24 - r_0^2r_c^3/12, \\
c_1 	& = 	&	r_0^2r_c^2 /4,\\
c_2 	& =	&	- r_0r_c(r_0 + r_c)/4, \\
c_3 	& = 	&	(r_0^2 + 4r_0 r_c + r_c^2)/12,\\
c_4 	& =	& 	-(r_0 + r_c) / 8,\\
c_5 	& =	& 1/20
\end{eqnarray}
The normalization is 
\begin{equation}
[w] = 4\pi \left[ r_0^3/3 + d\inv \sum_{i=0}^5 \frac{c_i}{i+3}(r_c^{i+3} - r_0^{i+3}) \right]
\label{eq:norm_smooth}
\end{equation}
The second derivative of the smooth indicator function is continuous everywhere and, in particular, vanishes for both $r \le r_0$ and for $r \ge \rc$.

\subsection{Sphere Indicator}
\label{SubSec-w-sphere}
We have also implemented a weighting function that describes the overlap between two spheres.
Consider a large sphere of radius $X$ and a small sphere of radius $x$, whose centers are separated by a distance $r$. 
We define
\begin{eqnarray}
X  	& = & (r_0 + r_c)/2		\\
x 	& = & (r_c - r_0)/2	.
\end{eqnarray}
such that $r_0 = X - x$ and $\rc = X + x$.  
The fraction of the small sphere that overlaps with the large sphere is  
\begin{equation}
\label{eq:sphere}
w(r) = 
\begin{cases}
	1 & r \leq r_0 \\
	c_{3} r^3 + c_1r + c_0 + c_{-1}/r & r_0 \leq r \leq r_c \\
	0 & \rc \leq r
\end{cases}
\end{equation}
where
\begin{eqnarray}
\label{eq-sphere-c}
c_{-1} 	& = &  -3(X^4 + x^4)/(16x^3) + 3X^2/(8x) \\
c_0 		& = &  (X^3 + x^3)/(2x^3) \\
c_1 		& = & -3(X^2 + x^2)/(8x^3) \\
c_3 		& = & 1/(16x^3)
\end{eqnarray}
The normalization is 
\begin{equation}
[w] 
	= 
		4\pi 
			\left[ 
				\frac{1}{3} r_0^3	
				+ 
				\sum_{i \in \{-1, 0, 1, 3\}} \left( \frac{c_i}{i+3} \right) (r_c^{i+3} - r_0^{i+3})
			\right]
\label{eq:norm_sphere}
\end{equation}
The first derivative of the sphere weighting function is continuous everywhere, while its second derivative is discontinuous at both $r_0$ and $\rc$.

\section{Hyperparameters for test calculations}
\label{sec:testparams}
\newcommand{\trim}{\epsilon_{\rm trim }}
\newcommand{\xlo}{x_{\zt; \rm lo}}
\newcommand{\xhi}{x_{\zt; \rm hi}}
\newcommand{\epsvd}{\epsilon_{\rm SVD}}
\newcommand{\bv}{{\mathbf{v}}}

BOCS employs several hyperparameters when performing variational force-matching.
As described in the main text, in addition to an input high resolution trajectory, the user must specify the type of interaction potentials, $\Uzt(x)$, that are to be included in the approximate CG potential, $U(\bR)$. 
(The same considerations also apply to the SG coefficient function, $\Udel$, but we describe interaction potentials for simplicity.)
Each interaction potential is a function of a mechanical degree of freedom, $\psi_\zt$, which may be, e.g., an angle or a type of local density.
The user must specify the domain, $[\xlo, \xhi]$, over which the corresponding force function, $\Fzt(x)$, should be determined. 
The lower, $\xlo$, and upper, $\xhi$, bounds of this domain are typically specified by the minimum and maximum values, respectively, of $\psi_\zt$ that are sampled by the input high resolution trajectory.
The user must also specify the type of basis functions (e.g., linear or cubic spline functions) employed to represent $\Fzt(x)$, as well as the width, $\dd x_\zt $,  of the support, $I_{\zt d}$, for each basis function, $\fzd(x)$.
In addition to these hyperparameters that define the basis functions for the approximate CG potential, BOCS also employs two additional hyperparameters, $\trim$ and $\epsvd$, to determine the coefficients for these basis functions.

The BOCS $\trim$ parameter determines a threshold for eliminating basis functions that have not been adequately characterized by the input high resolution trajectory.
Let $\De x_\zt = \xhi - \xlo$ be the width of the domain for the $\Fzt(x)$ force function.
The number of basis functions representing $\Fzt(x)$ is then $N_\zt = \De x_\zt / \dd x_\zt$.
When building the $\GDDp$ matrix and $\bD$ vector for the force-matching calculation, BOCS also determines a histogram, $h_{\zt d}$, that specifies the number of times the $\psi_\zt$ degree of freedom samples the interval, $I_{\zt d}$, associated with basis function $\fzd(x)$.
Let $H_\zt = \sum_d h_{\zt d}$ be the total number of times that the histogram is updated for the $\zt$ interaction.
If $h_{\zt d} < \trim \, H_\zt / N_\zt$, then BOCS eliminates the force field basis vector, $\cGzd$, associated with $\fzd(x)$ from the variational force-matching calculation, i.e., it trims the corresponding elements from the $G$ matrix and $b$ vector.

BOCS determines the optimal potential parameters, $\phi_*$, by solving the normal system of force-matching equations, $G \phi_* = b$, with the trimmed $G$ matrix and $b$ vector.
Let the eigenvalues of the trimmed $G$ matrix be ordered $\la_1 \ge \la_2 \ge \cdots \ge N_D$ and let $\bv_1, \bv_2, \ldots, \bv_{N_D}$ indicate the corresponding eigenvectors. 
Then $G\inv = \sum_{i=1}^{N_D} \bv_i \la_i\inv \bv_i\tr$. 
BOCS eliminates from this sum any terms associated with eigenvalues, $\la_i$, that are less than a user-specified threshold, $\epsvd \la_1$.

The manuscript reports the results of three tests that demonstrate the accuracy of the LD and SG potentials that we have implemented into BOCS version 5.0 and the PKG-BOCS package in LAMMPS.
Below we report the hyperparameters employed in the variational force-matching calculations for each test.
We report parameters in BOCS units, i.e., distances are reported in nm, densities are reported in ~nm$^{-3}$, and angles are reported in degrees.
When parameterizing CG models, we typically employ default values for these hyperparameters $\trim = 10^{-3}$ and  $\epsvd = 10^{-6}$.
However, in the reported test cases, we sometimes employed smaller values for these hyperparameters in order to more precisely recover the simulated force functions.

\newcommand{\rholo}{\rho_{\rm lo}}
\newcommand{\rhohi}{\rho_{\rm hi}}
\newcommand{\rlo}{r_{\rm lo}}
\newcommand{\rhi}{r_{\rm hi}}

\subsection{Test 1: 400 particles with linear LD potential}
Section 4.1.1 of the manuscript reports the results of force-matching calculations for a  system of 400 particles that interact via a linear LD potential. 
We simulated 5 variants of this system in which the local density was defined by the 5 weighting functions defined in section 1 above.
We performed two independent force-matching calculations for each of the 5 simulated variants.
\begin{enumerate}
\item
The first force-matching calculation represented $U$ with LD potentials.
This calculation employed a trimming parameter $\trim =  10^{-10}$ and SVD cutoff $\epsvd = 10^{-6}$.
In the case of the variant employing the Lucy weighting function, we employed a relatively coarse grid spacing, $\dd\rho$, and selected $\rholo$ to be somewhat larger than the minimum sampled local density in order to address sparse sampling of configurations with very low local density.
Table~\ref{tab:LDconstantrhoparams} reports the hyperparameters that define the basis functions for this force-matching calculation.

\begin{table}[h!]
	\centering
	\begin{tabular}{|c|c|c|c|}
		
		\hline
		$w$ 		& $[\rholo,\rhohi]$ 	& $\dd\rho$ 	& Spline Order \\
		\hline   DPD    & [2.3873241, 946.677] & 20 & 4 \\
		\hline   Lucy   & [100, 832.2593284] & 30 & 4 \\
		\hline   Shell  & [0.6266726, 250.2556252] & 20 & 4 \\
		\hline   smooth & [1.3369015, 533.9427756] & 20 & 4 \\
		\hline   sphere & [1.9098593, 758.2039449]& 20 & 4  \\
		\hline
	\end{tabular}
	\caption{
	Hyperparameters describing the force-matching calculations in section 4.1.1 that defined $U$ with LD potentials. The first column indicates the weighting function employed to define the local density for both the input simulation and the force-matching calculation. The second column indicates the support specified for the calculated LD force function, $\Frho(\rho)$. The third and fourth columns indicate the grid spacing and the order of the spline for the basis functions that represented $\Frho(\rho)$.  
}
	\label{tab:LDconstantrhoparams}
\end{table}

\item 
The second force-matching calculation represented $U$ with pair potentials. 
This calculation employed the same hyperparameters for each of the 5 variants.
In particular, it employed a trimming parameter $\trim =  10^{-9}$ and SVD cutoff $\epsvd = 10^{-6}$.
Table~\ref{tab:LDconstantpairparams} presents the hyperparameters defining the basis functions that represented the calculated pair force, $\Ftwo(r)$.

\begin{table}[h!]
	\centering
	\begin{tabular}{|c|c|c|c|}
		
		\hline
		$w$ 		& $[\rlo,\rhi]$ 	& $\dd r$ 	& Spline Order \\
		\hline   *    & [0.001, 1.5] 		& 0.01 		& 4 \\
		\hline
	\end{tabular}
	\caption{
	Hyperparameters describing the force-matching calculations in section 4.1.1 that defined $U$ with pair potentials. The first column indicates that the same parameters were employed for all 5 variants. The second column indicates the support specified for the calculated pair force function, $\Ftwo(r)$. The third and fourth columns indicate the grid spacing and the order of the spline for the basis functions that represented $\Ftwo(r)$.  
	}
	\label{tab:LDconstantpairparams}
\end{table}

\end{enumerate}

\subsection{Test 2: Two particles with LD or SG potentials}
Section 4.1.2 of the manuscript reports the results of force-matching calculations for a  system of 2 particles that interacted via either  LD or  SG potentials. 

\paragraph{LD potential}
We performed 5 simulations of 2 particles that interacted with LD potentials. 
The five simulations corresponded to defining the local density with the 5 weighting functions defined in section 1 above.
We performed two independent force-matching calculations for each of the 5 simulations. 

\begin{enumerate}
\item
The first force-matching calculation represented the potential with LD potentials.
This calculation employed a trimming parameter $\trim =  10^{-11}$ and SVD cutoff $\epsvd = 10^{-11}$.
Table~\ref{tab:LD2rhoparams} reports the remaining hyperparameters that define the basis functions for this force-matching calculation.

\begin{table}[h!]
	\centering
	\begin{tabular}{|c|c|c|c|}
		
		\hline
		$w$ 		& $[\rholo,\rhohi]$ 	& $\dd\rho$ 	& Spline Order \\
		\hline   DPD    & [0.00, 1376.6822386] & 138 & 4 \\
		\hline   Lucy   & [0.00, 1208.8031735] & 121 & 4 \\
		\hline   Shell  & [0.00, 362.6577722] & 37 & 4 \\
		\hline   smooth & [0.00, 773.669330] & 78 & 4  \\
		\hline   sphere & [0.00, 1103.0995341] & 111 & 4 \\
		\hline
	\end{tabular}
	\caption{
	Hyperparameters describing the force-matching calculations in section 4.1.2 that defined $U$ with LD potentials. The first column indicates the weighting function employed to define the local density for both the input simulation and the force-matching calculation. The second column indicates the support specified for the calculated LD force function, $\Frho(\rho)$. The third and fourth columns indicate the grid spacing and the order of the spline for the basis functions that represented $\Frho(\rho)$.  
	}
	\label{tab:LD2rhoparams}
\end{table}

\item 
The second force-matching calculation represented $U$ with pair potentials. 
This calculation employed the same hyperparameters for each of the 5 variants.
In particular, it employed a trimming parameter $\trim =  10^{-13}$ and SVD cutoff $\epsvd = 10^{-6}$ for each variant.
Table~\ref{tab:LD2pairparams} presents the hyperparameters defining the basis functions that represented the calculated pair force, $\Ftwo(r)$.

\begin{table}[h!]
	\centering
	\begin{tabular}{|c|c|c|c|}
		
		\hline
		$w$ 		& $[\rlo,\rhi]$ 	& $\dd r$ 	& Spline Order \\
		\hline   All indicators & [0.0000, 0.2] & 0.0001 & 2 \\
		\hline
	\end{tabular}
	\caption{
	Hyperparameters describing the force-matching calculations in section 4.1.2 that defined $U$ with pair potentials. The first column indicates that the same parameters were employed for all 5 variants. The second column indicates the support specified for the calculated pair force function, $\Ftwo(r)$. The third and fourth columns indicate the grid spacing and the order of the spline for the basis functions that represented $\Ftwo(r)$.  
	}
	\label{tab:LD2pairparams}
\end{table}

\end{enumerate}

\FloatBarrier

\paragraph{Square Gradient Case}
We performed 5 simulations of 2 particles that interacted via  SG potentials. 
The five simulations corresponded to defining the local density with the 5 weighting functions defined in section 1 above.
We performed two independent force-matching calculations for each of the 5 simulations. 

\begin{enumerate}
\item
The first force-matching calculation represented the potential with SG potentials.
This calculation employed a trimming parameter $\trim =  10^{-11}$ and SVD cutoff $\epsvd = 10^{-11}$.
Table~\ref{tab:SG2rhoparams} reports the remaining hyperparameters that define the basis functions for this force-matching calculation.
We represented the SG coefficient function for each variant with a single linear spline function.

\begin{table}[h!]
	\centering
	\begin{tabular}{|c|c|c|c|}
		
		\hline
		$w$ 		& $[\rholo,\rhohi]$ 	& $\dd\rho$ 	& Spline Order \\
		\hline   DPD    & [0.00, 298.3718880] & 298.3718880 & 2\\
		\hline   Lucy   & [0.00, 259.6008068] & 259.6008068 & 2 \\
		\hline   Shell  & [0.00, 78.3340522] & 78.3340522 & 2 \\
		\hline   smooth & [0.00, 167.0996250] & 167.0996250 & 2 \\
		\hline   sphere & [0.00, 237.8856639] & 237.8856639 & 2 \\
		\hline
	\end{tabular}
	\caption{
	Hyperparameters describing the force-matching calculations in section 4.1.2 that defined $U$ with SG potentials. The first column indicates the weighting function employed to define the local density for both the input simulation and the force-matching calculation. The second column indicates the support specified for the calculated SG coefficient function, $\Udel(\rho)$. The third and fourth columns indicate the grid spacing and the order of the spline for the basis functions that represented $\Udel(\rho)$.  
	}
	\label{tab:SG2rhoparams}
\end{table}

\item
The second force-matching calculation represented $U$ with pair potentials. 
This calculation employed a trimming parameter $\trim =  10^{-13}$ and SVD cutoff $\epsvd = 10^{-6}$ for each variant.
Table~\ref{tab:SG2pairparams} presents the hyperparameters defining the basis functions that represented the calculated pair force, $\Ftwo(r)$.

\begin{table}[h!]
	\centering
	\begin{tabular}{|c|c|c|c|}
		
		\hline
		$w$ 		& $[\rlo,\rhi]$ 	& $\dd r$ 	& Spline Order \\
		\hline   DPD & [0.0000, 0.2] & 0.0005 & 4 \\	
		\hline   Lucy & [0.0000, 0.2] & 0.002 & 4 \\
		\hline   Shell & [0.0000, 0.2] & 0.0005 & 4 \\
		\hline   smooth & [0.0000, 0.2] & 0.0005 & 4 \\
		\hline   sphere & [0.0000, 0.2] & 0.0005 & 4 \\
		\hline
	\end{tabular}
	\caption{
		Hyperparameters describing the force-matching calculations in section 4.1.2 that defined $U$ with pair potentials. The first column indicates the weighting function employed to define the local density for both the input simulation and the force-matching calculation. The second column indicates the support specified for the calculated pair force function, $\Ftwo(r)$. The third and fourth columns indicate the grid spacing and the order of the spline for the basis functions that represented $\Ftwo(r)$.  
	}
	\label{tab:SG2pairparams}
\end{table}

\end{enumerate}

\subsection{Test 3: 125 molecules with LD and SG potentials}
Section 4.1.3 of the manuscript reports a force-matching calculation for a  system of 125 molecules that interacted with bonded potentials, as well as pair, LD, and SG potentials. 
The force-matching calculation determined the force function, $F_\zt$, for all interactions except the SG potential. 
In the case of the SG potential, the force-matching calculation determined the SG coefficient functions, $U_{\del \tau_2|\tau_1}$. 
Table~\ref{tab:complicatedparams} reports the hyperparameters that defined the basis functions for this calculation.
The force-matching calculation employed the default trimming and SVD parameters, $\trim =  10^{-3}$ and $\epsvd = 10^{-6}$.

\begin{table}[h!]
	\centering
	\begin{tabular}{|c|c|c|c|c|}
		
		\hline
		Interaction  & $\psi_\zt$ & $[\xlo,\xhi]$ & $\dd x$ & spline order	\\
		\hline   All nonbonded pair 
					& $r$ 	& [0.00, 1.5]	& 0.01 	& 4\\
		\hline  All X$|$A LD  
					& $\rho$ 	& [0.00, 6.0] 	& 0.01	 & 4 \\
		\hline  X$|$B, X$|$C, X$|$D LD  
					& $\rho$ 	& [0.00, 6.0] 	& 0.1	 	& 4 \\
		\hline  All SG  
					& $\rho$ 	& [0.00, 6.0] 	& 0.1 	& 4 \\
		\hline  All Bonds  
					& $r_b$ 	& [0.00,0.45] 	& 0.0001	& 2 \\
		\hline  Angles ABC, BCD, ABA, DCD  
					& $\phi$ 	& [0.00, 180] 	& 1	 	& 2 \\
		\hline  Angle CDA  
					& $\phi$ 	& [0.00, 180] 	& 1	 	& 4 \\
		\hline  All Dihedrals  
					& $\psi$ 	& [-180, 180] 	& 3		& 4 \\
		\hline
	\end{tabular}
	\caption{
	Hyperparameters defining the basis functions for the force-matching calculation reported in  section 4.1.3. The first column indicates the type of interaction, while the second column indicates the corresponding mechanical degree of freedom. The third column indicates the support specified for the calculated force function, $F_\zt(x)$, or SG coefficient function, $U_{\del \tau_2|\tau_1}(\rho)$. The fourth and fifth columns indicate the grid spacing and the order of the spline for the basis functions that represented each of these functions.  
	}
	\label{tab:complicatedparams}
\end{table}

\section{Additional analysis of test 3 simulation}
\label{sec:MolecularAnalysis}
Section 4.1.3 of the manuscript reports a simulation and force-matching calculation for a  system of 125 molecules that interacted with bonded potentials, as well as pair, LD, and SG potentials. 
These molecules consisted of 4 types of sites $\tau = $ A, B, C, and D, with B and C sites at the interior of the molecule, while A and D sites occur at the surface.
Figure~6 of the manuscript reports the simulated force functions and SG coefficient functions.
Importantly, all site pairs interact via the same pair potential, while the LD and SG potentials acting on each site, $I$, are specified by the type, $\tau_I$, of this site. 
The LD and SG potentials become stronger in the order A $<$ B $<$ C $<$ D, i.e, they are weakest for sites of type A and strongest for sites of type D.
Moreover, the LD and SG potentials only generate finite forces when the corresponding local density falls in the range 1.63~\invnm $ <  \rhoL < $ 5.35~\invnm~and 1.91~\invnm $ <  \rhoL < $ 5.44~\invnm, respectively.

\begin{figure}[h!]
	\includegraphics{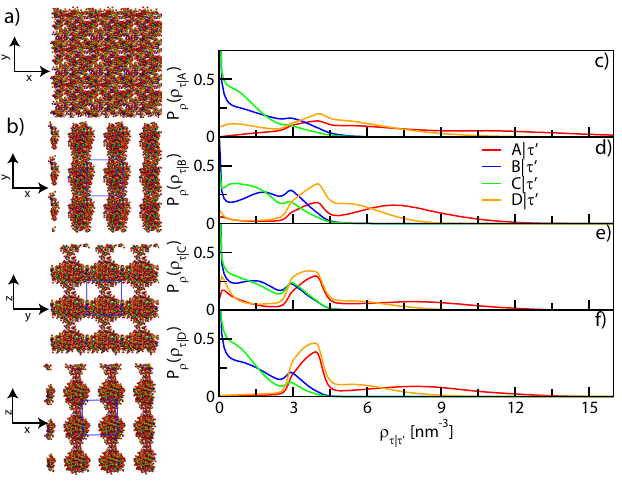}	
	\caption{
	Further analysis of the simulation from section 4.1.3 of the manuscript. 
	Panels and b report orthographic projections of the system at $t = $ 0~ns and $t = $ 6~ns, respectively.
	Panels c--f report the simulated local density distributions as a function of $\rho_{\tau|I}$.
	Panels c, d, e, and f correspond to sites $I$ with $\tau' = \tau_I = $ A, B, C, and D, respectively.
	In each of these panels, red, blue, green, and orange curves correspond to $\tau = $ A, B, C, and D, respectively.
	}
	\label{fig:molecularLDPDFs}
	
\end{figure}

Figure~\ref{fig:molecularLDPDFs} further characterizes the morphology generated by simulations with this model.
Panels a and b, which are reproduced from Fig.~7 of the manuscript, illustrate the initial condition of this simulation and the structure generated after 6~ns of simulation.
Panels c--f report the distribution of each type of local density, $\rho_{\tau|I}$, around each type of CG site.
As noted in the main text, the local densities of B and C sites, $\rho_{B,C |I}$ rarely exceed 4.5~\invnm.
Additionally, we note that the peak in the LD distributions of A and D sites become increasingly sharp around a site, $I$, as the type, $\tau_I$, of site $I$ varies in the order order A $<$ B $<$ C $<$ D, which corresponds to the order of the LD and SG potentials.

\begin{figure}[h!]
	\includegraphics{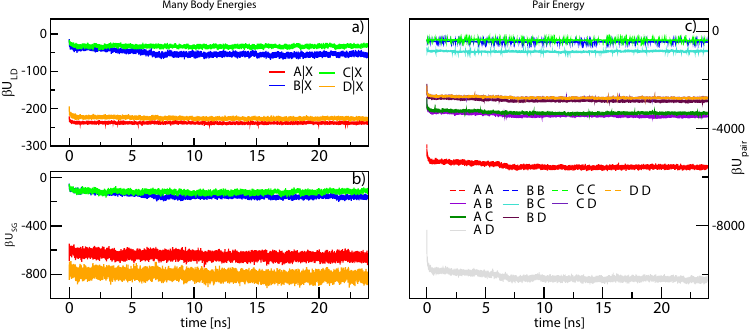}
	\caption{
	Analysis of the time evolution of the total nonbonded potential. 
	Panels a, b, and c report contributions from LD, SG, and pair potentials, respectively.
	The LD and SG potentials are decomposed into contributions from each site type, $\tau$,  by summing the contributions of this type to the LD and SG potentials for all of the particles. 
	}
	\label{fig:molecularenergies}
\end{figure}

Figure~\ref{fig:molecularenergies} reports time traces for various contributions to the total nonbonded potential.
Panels a and b report the contributions from LD and SG potentials.
In particular, we decompose the total LD and SG potentials into contributions from each type of local density around all of the particles, e.g., the A$|$X LD contribution, $U_{\rho \rm A|X}$,  is defined as the sum of contributions from A local densities around all $I = 1, \ldots, N$ particles, i.e., $U_{\rho \rm A|X}= \sum_{I=1}^N U_{\rho A|\tau_I}(\rho_{A|I})$.
Panel c reports the contributions from each nonbonded pair potential. 
The time traces demonstrate that variations in the pair potentials and, in particular, the AD pair potential dominate the change in the total nonbonded potential during the simulation.

\begin{figure}[h!]
	\includegraphics{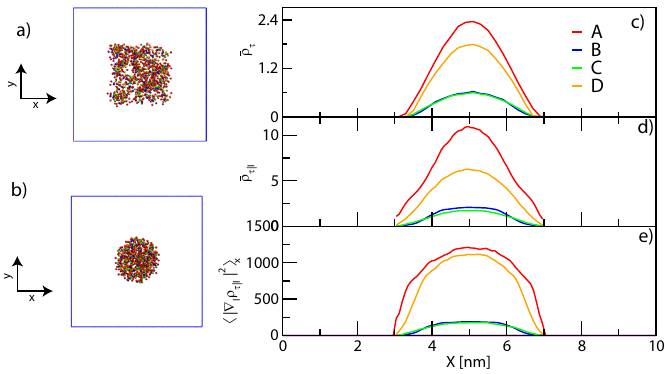}
	\caption{
	Analysis of simulated morphology.
Panel a presents a visualization of the initial configuration, while panel b presents a visualization of the configuration after 6~ns of simulation. 
Panels c -- e characterize the average density of $\tau = $ A, B, C, and D sites as a function of the x-coordinate in the simulation box. 
Specifically, 
panel c presents the average total density, $\brho_{\tau}(x)$, of $\tau$ sites; 
panel d presents the average local density, $\brho_{\tau|I}(x)$, of $\tau$ sites around each site, $I$, of type A; 
and 
panel e presents the average square gradient of the $\tau$ local density, 
$\llangle |\dI \rho_{\tau|I} |^2 \rrangle_x$, 
around each site, $I$, of type A. 
Panels c, d, and e report densities and square gradients in units of c) \invnm; d) \invnm; and  e)$\text{nm}^{-8}$.
	}
	\label{fig:dropsimulation}
\end{figure}

We investigated the influence of periodic boundary conditions upon the simulated morphology by simulating the same system in a larger simulation box.
Specifically, we followed the same protocols described in section~3.2.3 of the manuscript, except that we created the initial condition by randomly inserting the 125 molecules into a cubic region with sides of length $L=4$~nm in the center of a larger periodic cube with sides of length $L=10$~nm.
In this case the system rapidly evolved to form an approximately spherical drop.
Figure~\ref{fig:dropsimulation} analyzes this simulation in analogy to Fig.~7 of the manuscript.

\section{Data Availability}
The data and analysis for this study\cite{DataCommonsBocs2026} are available at the Penn State Data Commons repository\cite{DataCommons} repository with 	doi:10.26208/PW48-SA96.

\bibliography{WorkingBibFile}